\documentclass[sigconf]{acmart}
\setcopyright{none}
\renewcommand\footnotetextcopyrightpermission[1]{}
\acmConference{}{}{}
\acmBooktitle{}
\acmPrice{}
\acmISBN{}
\acmDOI{}
\AtBeginDocument{%
  }

\usepackage{xcolor}
\usepackage[normalem]{ulem}
\usepackage{subcaption}
\usepackage{float}
\usepackage{array}
\usepackage{caption}
\usepackage{tabularx}
\usepackage{multirow}
\usepackage{enumitem}
\usepackage{longtable}
\usepackage{booktabs}
\usepackage{threeparttable}
\usepackage{makecell}


\begin{document}

\title{``We Are the Dirty Hands'': Reimagining an Ambient Intelligence Care System with Professional Caregivers}

\author{Andrea Green}
\orcid{0009-0004-6973-4552}
\affiliation{%
  \institution{Stanford University}
  \city{Stanford}
  \state{CA}
  \country{USA}}

\author{Jane L. E}
\orcid{0000-0001-6868-7732}
\affiliation{%
  \institution{National University of Singapore}
  \city{Singapore}
  \country{Singapore}}

\author{Sneha Jain}
\orcid{0000-0001-7981-2754}
\affiliation{%
  \institution{University of Calgary}
  \city{Calgary}
  \state{AB}
  \country{Canada}}

\author{Parker Ruth}
\orcid{0000-0003-1030-4916}
\affiliation{%
  \institution{Stanford University}
  \city{Stanford}
  \state{CA}
  \country{USA}}

\author{Matthew Jörke}
\orcid{0000-0003-2972-462X}
\affiliation{%
  \institution{Stanford University}
  \city{Stanford}
  \state{CA}
  \country{USA}}

\author{Sarah L. Billington}
\orcid{0000-0003-3796-587X}
\affiliation{%
  \institution{Stanford University}
  \city{Stanford}
  \state{CA}
  \country{USA}}

\author{James A. Landay}
\orcid{0000-0003-1520-8894}
\affiliation{%
  \institution{Stanford University, Institute for Human-Centered Artificial Intelligence}
  \city{Stanford}
  \state{CA}
  \country{USA}}

\author{Andrea Cuadra}
\orcid{0000-0002-1845-0240}
\affiliation{%
  \institution{Olin College of Engineering}
  \city{Needham}
  \state{MA}
  \country{USA}}

\renewcommand{\shortauthors}{Andrea Green et al.}

\begin{abstract} 
While ambient assisted living (AAL) sensing systems offer promise for aging in place, their impacts on professional caregivers are overlooked. We conducted two collective speculation workshops ($N$=24) with personal health aides (PHAs, $n$=18) and certified nursing assistants (CNAs, $n$=6) to examine the ethical implications and value tensions of a \textit{care system with ambient intelligence (CSAI)}, our conceptualization of an AAL. Participants anticipated different implications across care roles and settings. PHAs anticipated additional responsibilities from \textit{CSAI}, including learning how it operates and supporting older adults in its use. CNAs, who had greater access to workplace resources and protections, similarly noted that \textit{CSAI} could create additional work. Participants also identified tensions around visibility, as sensing could support caregivers by documenting difficult working conditions while simultaneously making their behavior more vulnerable to misinterpretation. Through their modifications and surfaced discussions, we propose design guidelines for AAL systems, such as integrating context-aware sensing. 
\end{abstract}


\maketitle

\section{Introduction}
As care increasingly moves into the home, technologies that support aging in place introduce sensing into an environment that is both an older adult's home and a professional caregiver's workplace. Aging in place, or enabling older adults to remain in their homes as they age~\cite{NIA2023AgingInPlace}, has become an increasingly important goal~\cite{ratnayake2022aging}. The need for aging in place is in part due to the increasing older adult population~\cite{who2024ageing} and the anticipated strain it will put on the healthcare system~\cite{ratnayake2022aging}. In addition, most older adults prefer to remain in their homes for as long as possible~\cite{AARP2021WhereWeLive}, seeing improvements in independence and wellbeing when doing so~\cite{engineer2018designing}. To support this goal, researchers have explored technological methods such as ambient assisted living (AAL) sensing systems that combine ambient sensing and artificial intelligence (AI) to monitor personal and environmental signals, identify wellbeing and behavioral changes, and inform personalized care through recommendations and interventions~\cite{vimarlund2021ambient}. Such systems comprise configurations of sensors (e.g., cameras, motion sensors, and floor sensors) and devices (e.g., voice assistants) placed throughout the home and running continuously in the background, as well as applications that deliver AI-powered recommendations/interventions, which older adults and caregivers may have to interact with to mitigate situations that impact older adults' health (e.g., falls that lead to hospitalization or significant debilitating injuries)~\cite{ariani2016challenges,munoz2011design}.

As more older adults seek to receive care at home, facilitating aging in place will depend on professional caregivers. In particular, demand is expected to grow for personal health aides (PHAs)~\cite{dill2023will}, professional home care workers who assist older adults with activities of daily living~\cite{campbell2021caring}. Introducing AAL sensing systems into older adults' homes means they could potentially provide caregivers with information to perform their work efficiently but could also use the cameras and sensors to monitor their labor through systems they did not choose or design. These concerns are particularly important for PHAs as they work in private homes with limited institutional oversight and often precarious working conditions~\cite{campbell2021caring,ming2023go,karlsson2019home}. Introducing continuous sensing technology into homes with caregivers could serve as a form of worker surveillance~\cite{chowdhary2023can,alotaibi2016information,das2024sensible} resulting in data being used against their interests~\cite{kawakami2023sensing}. Additionally, AAL design work centers older adults' needs, values, and preferences, despite caregivers being parties potentially impacted by the systems. Therefore, designers should understand how AAL will impact caregivers' work and wellbeing to design for joint consideration of older adults and caregivers is crucial.

While there has been much work on how workplace sensing impacts workers~\cite{chowdhary2023can,alotaibi2016information,das2024sensible} and the use of technology and AI systems in care work~\cite{solano2025running,bartle2023machine,bartle2022second,solano2025eye,poon2021computer,ming2024wage,tseng2020we},
we know little about how caregivers anticipate the individual, community, and societal impacts of ambient systems that continuously observe and interpret activity within the home and what it means for professional care work. There is also a gap in understanding how to design AAL in a way that also empowers professional caregivers.

In this work, we speculate with professional caregivers about the potential ethical implications of integrating AAL sensing systems into older adults' homes. We use collective speculation~\cite{haghighi2023workshop}, a participatory method that enables impacted parties to anticipate the implications of emerging technologies and propose modifications that better reflect their needs and values. This approach allows us to develop design guidance grounded in caregivers' expertise before such systems are widely deployed with the aim of mitigating harms and ethical risks that AAL systems can have on professional caregivers. We conducted two collective speculation workshops ($N=24$) with PHAs and certified nursing assistants (CNAs) to examine the ethical implications of AAL sensing systems. PHAs ($n=18$) provided a community perspective as caregivers who would primarily encounter AAL in home settings, while CNAs ($n=6$), who work in skilled facilities (e.g., assisted living and nursing homes), provided perspectives on broader societal impacts and PHA career advancement\footnote{Becoming a CNA represents a common next career step for PHAs~\cite{kp2024}}. For this study, we conceptualize AAL as a \textit{\textbf{care system with ambient intelligence (CSAI)}}, a speculative concept that aims to support both older adults' care needs and caregivers' informational and career-development needs. We expand the definition to include caregiver support, as we anticipate this becoming a future capability of AAL, given the growing body of work on ambient technologies and AI-enabled caregiver support~\cite{bartle2022second,bartle2023machine,solano2025running}. Instead of a deployed system, \textit{CSAI} serves as an early prototype through which participants can consider the positive and negative implications of \textit{CSAI}, surface value tensions, and reimagine \textit{CSAI}. 

We explored the following research questions:

\begin{enumerate}[leftmargin=5em, itemsep=0.25em]
\item[\textbf{RQ1:}] \textbf{Ethical implications.} What ethical implications do PHAs and CNAs anticipate from \textit{CSAI}?

\item[\textbf{RQ2:}] \textbf{Value tensions.} What value tensions do PHAs and CNAs identify regarding \textit{CSAI}?

\item[\textbf{RQ3:}] \textbf{Design guidelines.} How can \textit{CSAI} be designed to address these ethical implications and value tensions?
\end{enumerate}

\section{Contribution}
With this work, we make \textbf{three contributions} to HCI research on ambient systems for aging in place, care work, and caregiver wellbeing. 

\begin{enumerate}
    \item We provide empirical findings on the ethical implications PHAs and CNAs anticipate with \textit{CSAI} in older adults' homes. We show how differences in caregivers' training, resources, and authority may leave them responsible for issues with \textit{CSAI} that they may not be adequately supported or empowered to address.

    \item We identify value tensions within and between participant groups. PHAs were particularly concerned that \textit{CSAI} could reduce their hours or threaten their jobs, while CNAs were less concerned about replacement and instead emphasized how existing institutional protocols and responsibilities could constrain the system's use.

    \item We propose caregiver-informed design guidelines for AAL sensing systems, including adapting systems to distinct care roles and settings and providing caregivers control over how information about their work is interpreted.
\end{enumerate}

\section{Related Work}
We first position our work within research on ambient systems for aging in place. We examine ambient technologies and AAL, including how these systems have been designed around older adults and their care. Then, we examine professional care work and care relationships, technologies that support professional caregiving, and the involvement of professional caregivers in the design of technologies for older adults.

\subsection{Ambient Systems for Aging in Place}
\paragraph{\textbf{Ambient Technologies and Intelligence}}
Ambient technologies are computing devices embedded into everyday environments that can sense and collect information in the background with little or no direct input from people~\cite{cicirelli2021ambient}. These technologies can include wearable devices that monitor health and activity, smart infrastructure such as thermostats and lighting systems, and environmental sensors that capture motion and occupancy~\cite{cicirelli2021ambient}. Recent HCI work has also explored ubiquitous computing devices, such as voice assistants, as ambient technologies to capture visual and audio from their environments~\cite{cuadra2023designing,pathirana2026older}. These capabilities have expanded opportunities to use ambient technologies for health applications in the built environment, where information about people’s activities, behaviors, and environments can be used to monitor changes in their health and wellbeing and respond~\cite{acampora2013survey}.

\paragraph{\textbf{Ambient Assisted Living and Older Adult Care}}
Ambient Assisted Living (AAL) applies ambient technologies and intelligence to support older adults aging in place~\cite{jovanovic2022ambient,cicirelli2021ambient,vimarlund2021ambient}. AAL sensing systems can continuously collect information about a person and their environment and use these data to identify changes that may indicate a need for care support or health intervention~\cite{jovanovic2022ambient}. For example, floor and motion sensors can capture changes in gait, movement, and activity associated with fall risk and use these data to inform strategies for reducing that risk~\cite{muheidat2020home,camp2020technology}. Bringing these monitoring capabilities into the home aims to help older adults remain safely in familiar environments while maintaining greater independence, autonomy, and agency~\cite{jovanovic2022ambient}. Given these aims, AAL design research has historically centered older adults' needs, values, and preferences~\cite{choukou2021attitude,van2011ageing,reeder2014assessing}. This work has raised concerns about privacy~\cite{choukou2021attitude,van2011ageing}, system complexity~\cite{reeder2014assessing}, and control over what information is collected and shared~\cite{choukou2021attitude}. These conversations have expanded to HCI to consider how AAL may operate within home care and how the information these systems produce is interpreted. For example, interventions recommended by intelligent systems raise questions of liability when recommendations are inaccurate~\cite{green2025black}. Research with low-income and Black older adults has also shown how ambient data may be interpreted without sufficient context. For example, data revealing drug use or diagnoses of sexually transmitted diseases, for example, may lead to judgment~\cite{so2024they}, while health interventions that do not account for users' racial backgrounds may not align with their needs~\cite{green2025black}. 

As AAL has become increasingly situated within the broader network of home care, research has also considered the people who use the information these systems produce to support older adults. These users have largely been conceptualized as informal caregivers, unpaid caregivers who are typically an older adult's family members~\cite{schulz2010informal}. Informal caregivers may view AAL as a support tool to maintain awareness of a loved one's safety while reducing some of the uncertainty and burden associated with care that they are largely untrained to provide ~\cite{jaschinski2019listening}. Thus, although AAL design has commonly centered older adults' adoption, use, and care, older adults and informal caregivers may not be the only people who will interact with these systems in the future.

\subsection{Professional Caregiving, Care Relationships, and Technology}
\paragraph{\textbf{Professional Care Work and Care Relationships}}
Aging in place does not mean aging without assistance. As older adults require support with everyday activities, they may rely on professional caregivers who provide care in their homes~\cite{campbell2021caring,stone2019future}. Personal health aides (PHAs) include personal care aides and home health aides, who support older adults in their homes with activities of daily living, household tasks, and health monitoring, making their work an important part of the infrastructure that enables aging in place~\cite{stone2019future,ming2023go}. Yet this work is often performed under precarious conditions. Home care workers are disproportionately women and people of color, and nearly half live in low-income households~\cite{phi2019ushome}. Despite their contribution to aging in place, these workers experience low wages, unstable employment, limited training, and fewer institutional protections~\cite{campbell2021caring,phi2019ushome}. These conditions are further shaped by the home as a workplace. Unlike the caregivers who work in skilled facilities, PHAs often provide care alone and with limited supervision in environments that were not designed for care work~\cite{campbell2021caring,hittle2016complexity}. Home caregivers may encounter physical aggression, harassment, racialized remarks, and occupational risks associated with confined or cluttered work environments~\cite{lucien2024violence,karlsson2019home,hittle2016complexity,markkanen2017safety}. 

Caregivers' responsibilities also extend beyond formal care tasks. Caregivers build trust, provide emotional support, manage family expectations, and respond to needs outside of formal care plans---forms of relational and ``invisible work'' that are necessary for care but often go unrecognized and unpaid~\cite{ming2023go,campbell2021caring}. Therefore, professional caregiving is shaped by assigned care tasks and the physical and relational conditions of the home environment. These conditions become important as AAL instruments these environments. Ambient technologies and intelligence may provide information that supports care, but they can also capture aspects of caregivers' work~\cite{ming2023go,solano2025running}. Additionally, professional caregivers may be expected to provide the care needed to mitigate risks identified by these systems. Unlike informal caregivers, how they respond to these risks may become part of how their work performance is evaluated. Yet, professional caregivers have received comparatively less attention in the design of AAL~\cite{percy2024user,milella2023artificial}. 

\paragraph{\textbf{Technologies for Professional Care Work}}
HCI research has explored technologies supporting professional caregivers' access to care information, coordination, documentation, training, and workplace resources~\cite{ming2023go,bartle2022second,solano2025eye,poon2021computer,ming2024wage,tseng2020we}. Such resources can support job satisfaction and opportunities for professional development~\cite{campbell2021caring,hermansen2023immigrant}. At the same time, HCI research has shown that technologies introduced into care settings do not just support existing work; they can change what work is expected and documented~\cite{ming2023go}. However, this work has examined technologies that are meant as caregiver-facing tools~\cite{bartle2022second,bartle2023machine,ming2023go,poon2021computer}. Unlike caregiver-facing tools, AAL continuously senses the environment and can capture information about caregivers as they move through shared spaces, often without their direct control over what is sensed. Therefore, professional caregivers may become secondary users of information produced by AAL sensing systems while also becoming incidental subjects of their data collection. Workplace sensing can introduce concerns around surveillance, consent, and power when workers have limited control over what is collected or how those data are used~\cite{chowdhary2023can,alotaibi2016information,das2024sensible}. Additionally, AAL systems inherently center older adults and are not meant to directly advocate for care workers, resulting in tools that may not serve them. This work suggests that integrating ambient technologies into the home requires considering how their presence may impact caregivers' work and wellbeing.

\paragraph{\textbf{Professional Caregivers in the Design of Older Adult Technologies}}
Technologies designed only around older adults' needs and values can introduce different consequences for caregivers. For example,
\citet{so2024they} found that older adults envisioned using voice assistants to improve their relationships with caregivers by providing them with information to support care, but also as monitoring devices to mitigate caregiver theft. These design modifications both aim to improve the care relationship but have substantially different risks for the caregiver. Therefore, understanding the risks to caregivers is essential in early design stages. Involving professional caregivers in the design of technologies for older adult care has shown that their perspectives can surface needs that may be overlooked when technologies are designed primarily around older adults
~\cite{bartle2022second,bartle2023machine,ming2023go}. Therefore, we explore this intersection by examining the design of older adult-centered technology through caregivers' perspectives.

However, involving workers in technology design is particularly challenging when the workers have limited technical or design expertise~\cite{rosner2016out,harrington2019deconstructing}. Haghighi and J{\"o}rke et al.~\cite{haghighi2023workshop} developed collective speculation to support participants without design expertise in examining the human, community, and
societal implications of emerging technologies and creating speculative artifacts that modify or reimagine them~\cite{haghighi2023workshop}. These conversations are particularly relevant to AAL, where professional caregivers may be impacted by the systems they had little role in designing. Comparatively less is known about how professional caregivers themselves envision AAL sensing systems that may become part of their work environment. We build on these conversations by examining the ethical implications and and value tensions PHAs and CNAs anticipate within AAL and how they envision such systems better accounting for them.

\section{Method}
We conducted two IRB-approved 
collective speculation workshops with certified nursing assistants (CNAs) and personal health aides (PHAs) to examine their perspectives on the positive and negative implications of \textit{CSAI}, identify value tensions, and understand how they would modify \textit{CSAI} in response. We describe participants and recruitment, our adaptation and implementation of the collective speculation method, the design probes used to support speculation, data collection and analysis, and researcher positionality.

\subsection{Participants}
We conducted two workshops with 24 total participants (23F, 1M; mean age = 49, range = 27--67): six CNAs (C1--C6) participated in September 2024 and 18 PHAs (P1--P18) participated in November 2024. All PHA participants were Pacific Islanders\footnote{Asian and Pacific Islander communities are more likely to provide home care to older adults compared to the general population~\cite{montenegro2014caregiving}.}, and all CNAs except C4 reported previous experience working as PHAs. PHAs self-identified as either personal care aides ($n$=7), home health aides ($n$=7), or home caregivers ($n$=4). Table~\ref{tab:aggregate} summarizes participant demographics.

\begin{table}[htbp]
\centering
\small
\caption{Self-reported participant demographics aggregated ($N=24$)}
\label{tab:aggregate}
\begin{tabularx}{\columnwidth}{@{}p{0.35\columnwidth}X@{}}
\toprule
\textbf{Age} & Mean = 49, Mode = 50, Median = 49, SD = 10.4 \\
\midrule
\textbf{Gender} & Woman: 23, Man: 1 \\
\midrule
\textbf{Race} & Pacific Islander: 18, Asian: 1, Black: 1, White: 1, Declined to answer: 3 \\
\midrule
\textbf{Ethnicity} & Hispanic: 4 \\
\midrule
\textbf{Highest degree level} & Master's degree: 1, Bachelor's degree: 1, Some college: 6, High school diploma or GED: 13, Less than a high school diploma: 3 \\
\bottomrule
\end{tabularx}
\end{table}

We recruited CNAs through a continued care retirement community 
and PHAs through referrals from gerontology and home-based care physicians. Eligible participants had at least six months of caregiving experience and current or previous employment as a CNA or PHA. Because some PHAs had limited familiarity with computers, eligibility could also be assessed by email or phone. Participants received \$35 per hour and an additional \$100 for completing both workshop days.

Participants worked in groups of five or fewer (Table \ref{tab:groups}). The CNA workshop included two groups (C-G1 and C-G2), and the PHA workshop included four groups (P-G1--G4). Participants in C-G1 were primarily Spanish speakers with limited English proficiency and were grouped with a Spanish-speaking facilitator for translation. All PHA participants were bilingual in Tongan and English and completed workshop activities in English. Otherwise, there was no participant grouping criteria.

\begin{table*}[ht]
\centering
\caption{Participant groups ($N$=24)}
\label{tab:groups}
\small
\setlength{\tabcolsep}{3pt}
\renewcommand{\arraystretch}{1.05}

\begin{tabular*}{\textwidth}{
@{\extracolsep{\fill}}
l c c c c c c
}
\toprule
& \textbf{C-G1} & \textbf{C-G2} & \textbf{P-G1} &
\textbf{P-G2} & \textbf{P-G3} & \textbf{P-G4} \\
& ($n$=4) & ($n$=2) & ($n$=4) &
($n$=5) & ($n$=4) & ($n$=5) \\
\midrule

\textbf{Age, M (SD)}
& 47.5 (6.8)
& 34.5 (10.6)
& 45.0 (2.6)
& 61.8 (5.1)
& 44.3 (14.4)
& 50.4 (7.1) \\

\textbf{Gender} 
& W (4) 
& W (1); M (1) 
& W (4) 
& W (5) 
& W (4) 
& W (5) \\

\textbf{Race} 
& White (1); NR (3) 
& Black (1); Asian (1) 
& Pacific Islander (4) 
& Pacific Islander (5) 
& Pacific Islander (4) 
& Pacific Islander (5) \\

\textbf{Ethnicity} 
& Hispanic (4) 
& NR (2) 
& NR (4) 
& NR (5) 
& NR (4) 
& NR (5) \\

\bottomrule
\end{tabular*}

\vspace{2pt}
\begin{minipage}{\textwidth}
\footnotesize
\textit{Note.} Mean age is reported in years. 
W = woman; M = man; NR = not reported.
\end{minipage}

\end{table*}

\subsection{Collective Speculation Workshop}
We adapted the collective speculation workshop developed by Haghighi \& J{\"o}rke et al.~\cite{haghighi2023workshop}, a method for engaging people impacted by emerging technologies in anticipating ethical implications, exploring value tensions, and proposing design modifications without requiring prior design expertise. The workshop progresses through four phases: (1) Warm-up (introducing speculative design); (2)  Anticipation Round (identifying contexts, use cases, impacted direct and indirect parties, and positive and negative implications); (3) Implication Design Round (developing modifications to negative implications); and (4) Action Round (examining tensions between positive and negative implications through role-play and design). 

\paragraph{\textbf{Adaptations for Professional Caregivers}}
Collective speculation is intended to be adapted to the needs of the participant community~\cite{mack2022anticipate}. Through three pilot sessions with participants from Latin American, Asian, and Indigenous backgrounds, including Spanish-speaking participants, and interviews with a PHA, gerontologists, and a home care agency administrator, we identified three design goals: (DG1) reduce power dynamics and encourage balanced participation; (DG2) accommodate caregivers' schedules to reduce participation burden; and (DG3) support engagement with unfamiliar concepts without requiring design expertise. These goals reflected characteristics of the care context that could shape participation. In particular, hierarchical structures within healthcare settings (e.g., between home and facility workers) can make individuals less likely to voice concerns~\cite{rosen2018teamwork}, motivating us to minimize power differences within the workshops. Caregivers' demanding and often irregular work schedules~\cite{campbell2021caring,phi2019ushome} similarly motivated us to structure participation around their availability and reduce the additional burden of taking part.

We therefore divided the workshop into two approximately three-hour sessions, provided additional time for explanation and ideation, introduced a guided speculation exercise before the warm-up activity during the Warm-up phase, reduced the number of warm-up design constraints, assigned two participant moderators during the Action Round, and maintained groups across activities to preserve language support and established group dynamics. The CNA workshop occurred on weekday evenings and the PHA workshop on weekend mornings, with 20--30-minute breaks between major activities. Table~\ref{tab:modtable} summarizes the adaptations for each phase.

\begin{table*}[htbp!]
\caption{Adaptations to the collective speculation workshop for professional caregivers}
\label{tab:modtable}
\footnotesize
\setlength{\tabcolsep}{4pt}
\renewcommand{\arraystretch}{0.92}

\begin{tabularx}{\textwidth}{
    >{\raggedright\arraybackslash}p{0.10\textwidth}
    >{\raggedright\arraybackslash}p{0.20\textwidth}
    >{\raggedright\arraybackslash}p{0.25\textwidth}
    >{\raggedright\arraybackslash}X}
\toprule
\textbf{Phase} & \textbf{Original Method} & \textbf{Adaptation} & \textbf{Rationale} \\
\midrule

\multirow{3}{*}{Warm-up}
& 15--20 min
& Extended to 45--60 min
& \textbf{DG3.} More time to introduce speculative design and scaffold activities \\
\cmidrule{2-4}

& Immediate introduction of \textit{Toilet Buddy}
& Guided speculation before introduction of \textit{Toilet Buddy}
& \textbf{DG3.} Introduced the idea of ethical implications as benefits and concerns  \\
\cmidrule{2-4}

& Five prototype constraints
& Reduced to three
& \textbf{DG3.} Reduced complexity, fatigue, and and provided more time per idea \\
\midrule

Anticipation
& 45--60 min
& Extended to ${\sim}$75 min
& \textbf{DG3.} More time for explanation, brainstorming, and discussion \\
\midrule

Implication Design
& 45--60 min
& Extended to 60--75 min
& \textbf{DG3.} More time to interpret implications and develop modifications \\
\midrule

\multirow{4}{*}{Action}
& 45--60 min
& Extended to 60--75 min
& \textbf{DG3.} More time to explore relationships between implications \\
\cmidrule{2-4}

& One participant moderator per group
& Two moderators per group
& \textbf{DG1.} Distributed leadership and reduced reliance on one participant \\
\cmidrule{2-4}

& Participants shuffled between groups
& Groups remained intact
& \textbf{DG1, DG3.} Preserved language support and established group dynamics \\
\cmidrule{2-4}

& Implication cards drawn individually
& Cards selected collectively
& \textbf{DG2.} Reduced activity time while preserving group discussion \\
\bottomrule
\end{tabularx}

\vspace{1pt}
\scriptsize
\textit{Note.} DG1 = reduce power dynamics; DG2 = accommodate work schedules and participation burden; DG3 = support understanding of unfamiliar concepts.
\end{table*}

\subsection{Care System with Ambient Intelligence}
We developed the design probes through four interviews with caregiving and healthcare community members to understand caregiving roles and scenarios. We presented \textit{CSAI} as an intelligent home sensing system for simplicity, comprising (1) environmental sensors, wearables, and ubiquitous computing technologies in older adults' homes; (2) a tablet and voice assistant supporting older adults' wellbeing management; and (3) a caregiver-facing application supporting health monitoring, care interventions, communication, and career development and advancement. We used design probes to provide a shared representation of \textit{CSAI} while keeping the speculative system open to critique and modification. 

\paragraph{\textbf{Scenario}}
The scenario was intended to provide a description of the system's intended functionality. The scenario for \textit{CSAI} is as follows:

\begin{quote}
\textit{A new healthcare technology company has launched an intelligent home sensing system. The system uses sensors installed in the homes of older adults to improve their health and wellbeing by monitoring them and their environment. The system is connected to a voice assistant and an app installed on caregivers’ phones or tablets. The voice assistant helps older adults manage their health using sensor data and allows them to ask medical questions. The app supports caregivers’ career development by providing tutorials, assisting with documentation, and highlighting care insights from the data.}
\end{quote}

\begin{figure*}[!t]
    \centering
    \includegraphics[width=0.95\textwidth]{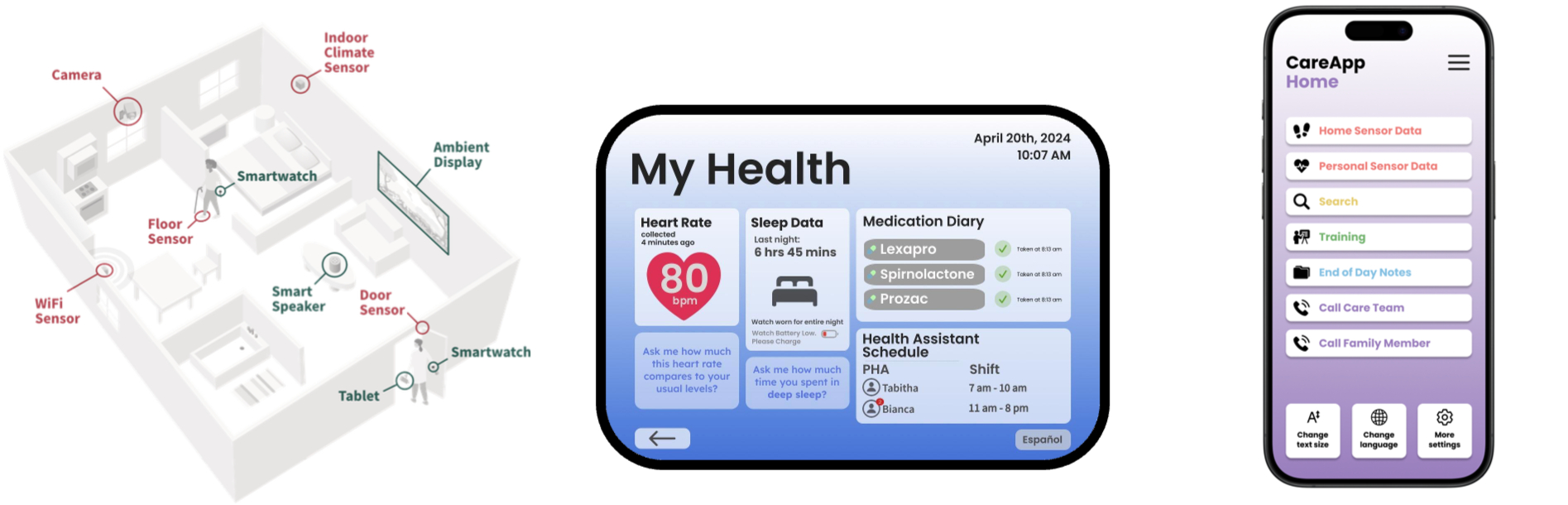}
    \caption{\textbf{Representation of \textit{CSAI}:} Overview presented to participants depicting (L--R): (1) an ecosystem of sensors, wearables, and ubiquitous computing devices integrated into the older adult's home to support both the older adult and caregiver; (2) a tablet interface for older adults supporting health management; and (3) a caregiver-facing application supporting patient health management, training and upskilling, and communication with the care team and family members.}
    \label{fig:systemoverview}
\end{figure*}

\paragraph{\textbf{Medium-fidelity prototypes}}
For medium-fidelity prototypes (Figure ~\ref{fig:systemoverview}), we used the a(x4) exercise~\cite{rothstein2000challenge}, Nielsen's usability heuristics~\cite{nielsen2005ten}, and expert review~\cite{vermeeren2010user} for developing the interfaces. Features were informed by prior work on accessibility~\cite{hou2022design,so2024they} and personalization features for older adults~\cite{green2025black}, caregiver-facing tools~\cite{bartle2023machine}, and caregiver training and career advancement~\cite{phi2019ushome}. We deliberately retained imperfect features for CNA exam questions and complex sensor visualizations to support critique rather than present \textit{CSAI} as an idealized system. 

\begin{figure}[htbp!]
    \centering
    \begin{subfigure}[b]{0.3\textwidth}
        \includegraphics[width=\textwidth]{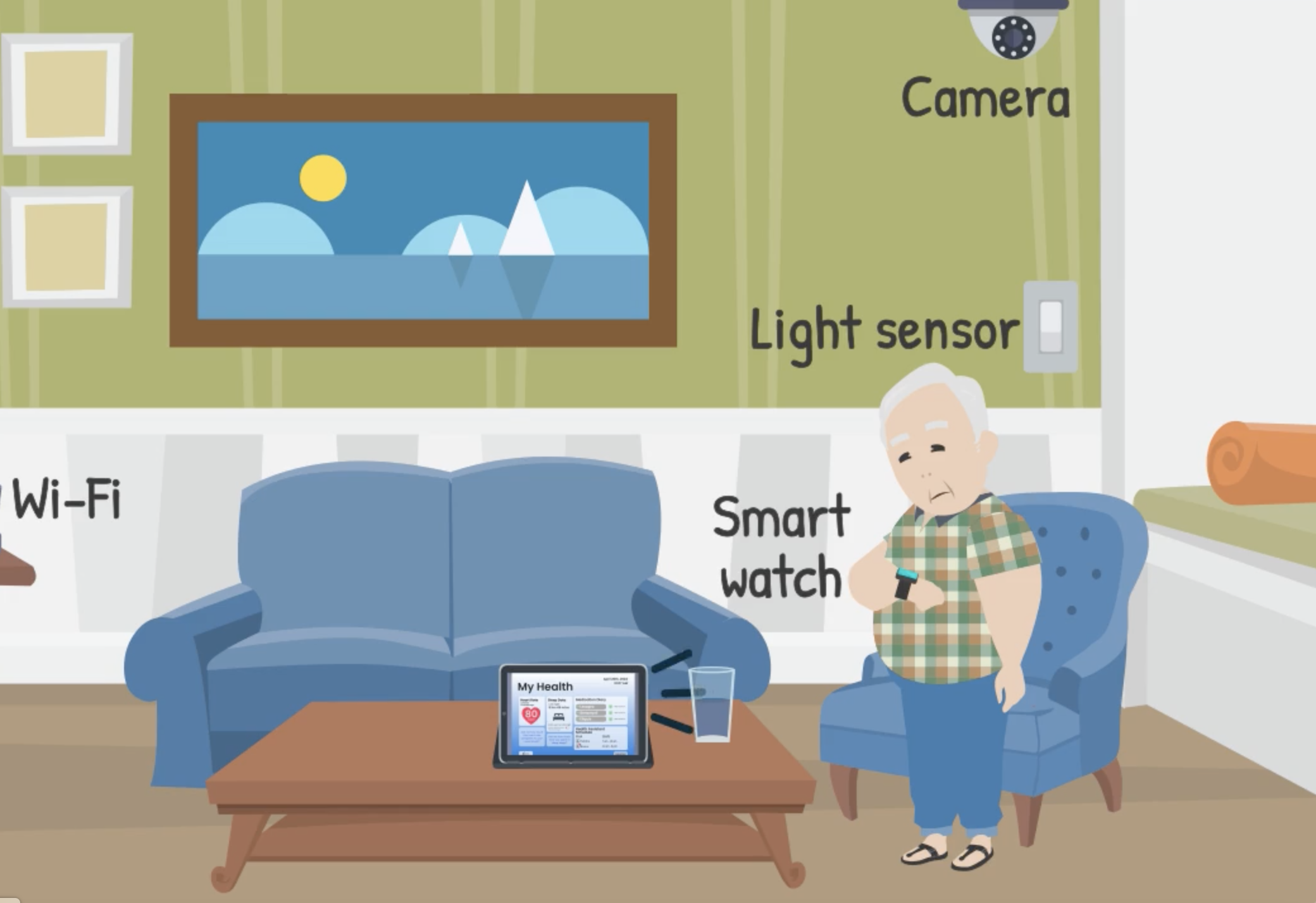}
        \caption{Sensors in the home}
        \label{fig:Video1}
    \end{subfigure}
    \hfill
    \begin{subfigure}[b]{0.3\textwidth}
        \includegraphics[width=\textwidth]{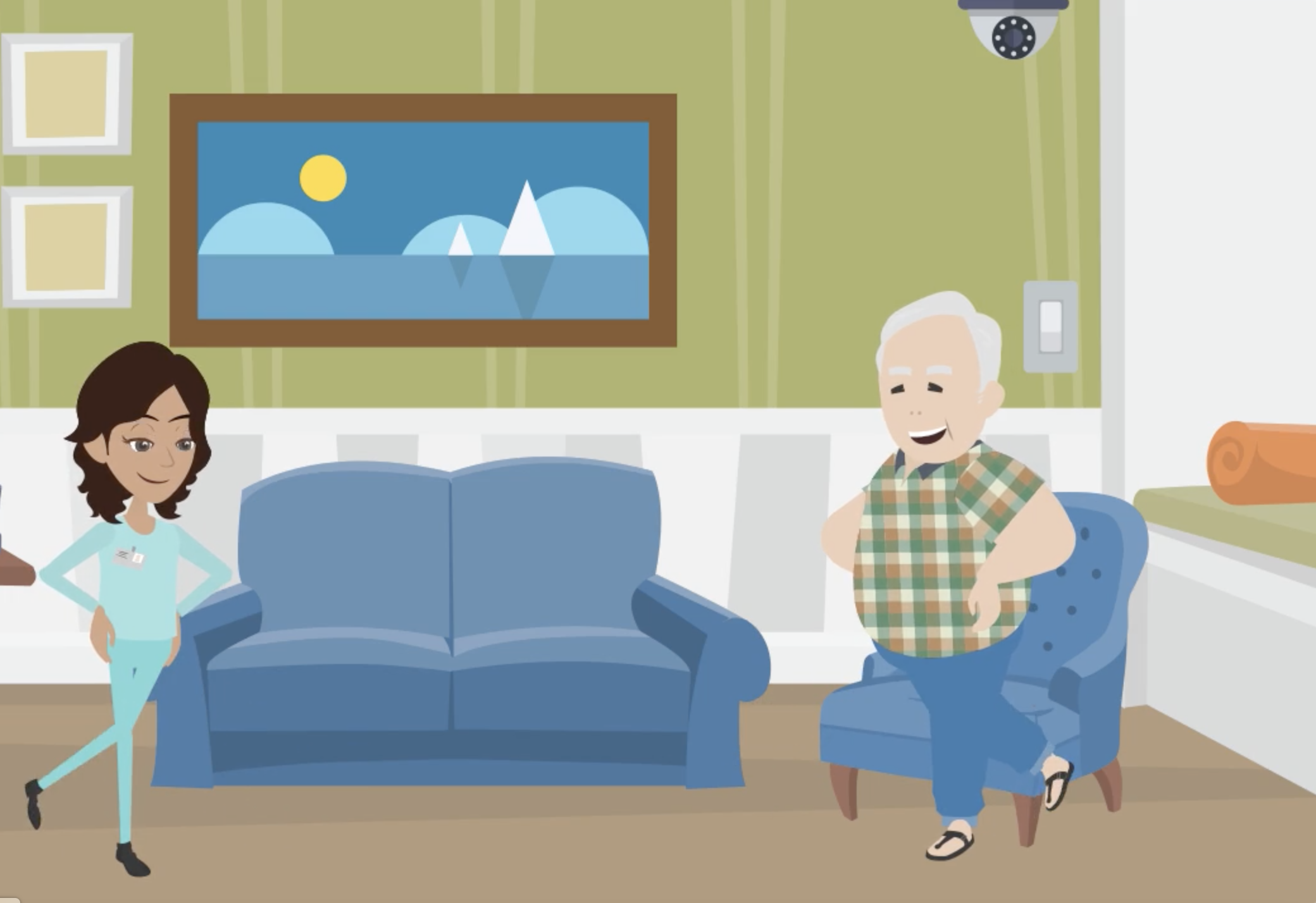}
        \caption{Collaborative balance exercise}
        \label{fig:Video2}
    \end{subfigure}
    \hfill
    \begin{subfigure}[b]{0.3\textwidth}
        \includegraphics[width=\textwidth]{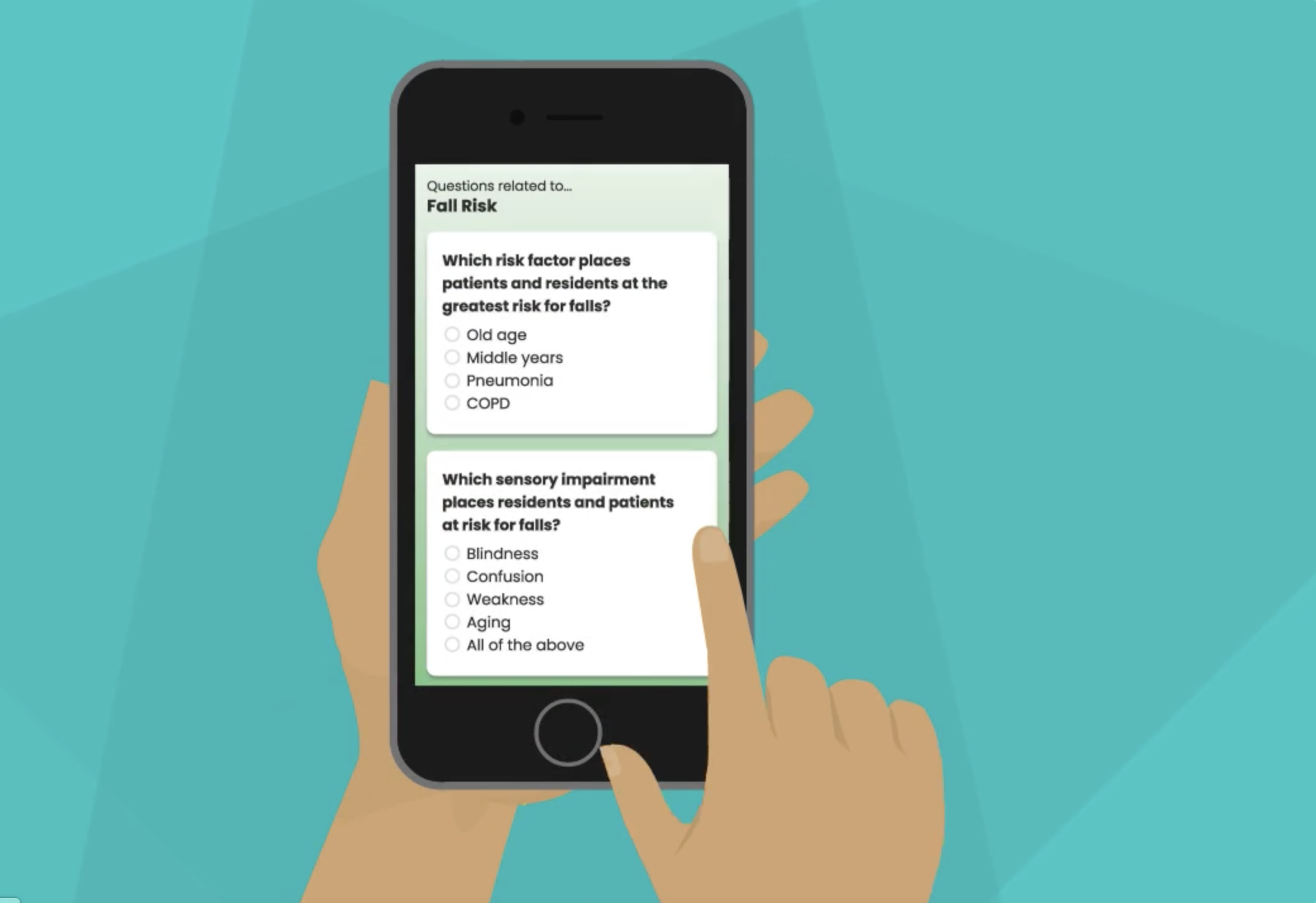}
        \caption{In-app caregiver support}
        \label{fig:Video3}
    \end{subfigure}
    \caption{Side-by-side screenshots of the concept video used to envision the functionality of \textit{CSAI}. The video was created using the Vyond platform \cite{Vyond_2025}.} \Description{Screenshots from a concept video showing the app functionality.} \label{fig:Video}
\end{figure}

\paragraph{\textbf{Concept video}}
A short concept video (Figure ~\ref{fig:Video}) illustrated how \textit{CSAI} could collect sensor data and generate an intervention, how a caregiver might facilitate that intervention, and how information generated through care activities could support career advancement. The racial background and gender of the caregiver and older adult depicted were based on an observation of a PHA and older adult at the continued care facility. 

The video scene presented to participants was as follows:

\begin{quote}
\textit{Mr. Jones interacts with Carol, an IVA embedded in his tablet, who introduces \textit{CSAI} and its use of ambient sensing to assess his wellbeing (Figure~\ref{fig:Video1}). When \textit{CSAI} predicts an increased fall risk, Carol advises him to sit and alerts his caregiver, Maria. Maria receives the alert while doing laundry and, after convincing, performs the suggested intervention with Mr. Jones (Figure~\ref{fig:Video2}). While Mr. Jones naps, Maria completes CNA exam practice questions generated from the care interaction (Figure~\ref{fig:Video3}).}
\end{quote}

\subsection{Workshop Procedure}
\begin{figure*}[!t]
    \centering
    \includegraphics[width=1.0\linewidth]{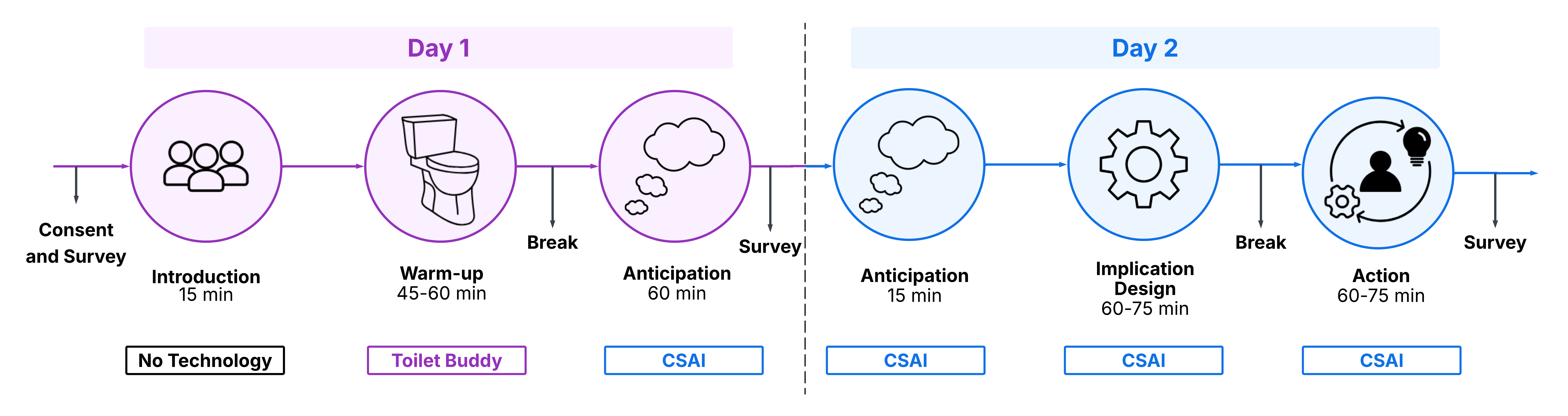}
    \caption{Overview of the workshop structure, including the phases and technology introduced during each.}
    \label{fig:procedure}
\end{figure*}

Figure~\ref{fig:procedure} summarizes the two-day workshop. Before participation, participants completed consent and demographic surveys with researcher assistance as needed. Workshops were conducted in a conference room with three to five facilitators: a lead facilitator responsible for instructions and timing and one facilitator assigned to each group. Each group received reference sheets of the technologies, guiding slides, blank note cards, Tarot Cards of Tech~\cite{TarotCardsOfTech}, and crafting materials. Full workshop materials are provided in supplemental materials. 

\paragraph{\textbf{Introduction (Day 1)}}
We set the tone for the workshop by establishing norms (e.g., encouraging open-mindedness and creating a judgment-free space for sharing ideas). Because caregivers may prioritize older adults' needs over their own~\cite{bartle2022second}, we explicitly encouraged participants to consider how the technologies could impact them as care workers. We also emphasized that responses were not being graded and encouraged participants to criticize/reject the technologies presented if they felt compelled to do so. We further introduced collective speculation and ethical implications.  

\paragraph{\textbf{Warm-up (Day 1)}}
Participants practiced speculative design using \textit{Toilet Buddy}, an intelligent toilet adapted from prior collective speculation workshops~\cite{haghighi2023workshop,so2024they}. Following a guided introduction to speculation, participants were introduced to \textit{Toilet Buddy} through a scenario and individually proposed modifications intended to preserve perceived benefits while mitigating harms under three ideation constraints: using magic, using technology from 100 years ago, and designing a modification that would result in them getting fired. These constraints encouraged broad ideation, attention to human needs instead of technology limits, and consideration of employment consequences. 

The scenario for \textit{Toilet Buddy} is as follows:
\begin{quote}
    \textit{To improve the health and wellbeing of older adults, Toilet Buddy will be installed in all of their bathrooms. Toilet Buddy is an intelligent toilet that cares about them and their wellbeing. Toilet Buddy monitors older adults’ nutrition information, vitamin deficiencies, and stress levels to provide customized nutrition plans and health recommendations to them and their caregivers over time. While it's meant for older adults, it may need to sense the caregiver’s information as well to determine who is using the toilet.}
\end{quote}

\paragraph{\textbf{Anticipation Round (Days 1 \& 2)}}
We introduced \textit{CSAI} via the design probes. Participants individually and collectively identified contexts, use cases, directly and indirectly affected parties, and positive and negative implications of \textit{CSAI}. These ideas were recorded on cards and subsequently used during the Action Round. Workshop \#1 repeated part of the activity on Day 2 to incorporate a newly joined participant (C5); Workshop \#2 continued the activity on Day 2 because of time constraints.

\paragraph{\textbf{Implication Design Round (Day 2).}}
Participants selected positive and negative implications generated during the Anticipation Round and individually developed three modifications to \textit{CSAI}: one involving a device, one a wearable, and one a screen-based intervention. Each modification was intended to communicate or protect against the selected negative implication and maintain the positive.

\paragraph{\textbf{Action Round (Day 2).}}
Each group selected two participant moderators, who paired one positive and one negative implication associated with the group's use case and facilitated role-playing around how these implications might coexist in care contexts. Participants then individually designed artifacts intended to preserve the positive implication while communicating or protecting against the negative implication. Designs could use magic but could not introduce new laws, assume an unhackable system, or address implications without a physical or digital artifact.

\paragraph{\textbf{Surveys (Days 1 \& 2)}}
At the end of each day, participants completed surveys containing Likert-scale and open-ended questions about their experience and confidence with the activities. The Day 2 exit survey additionally asked broadly about likes, dislikes, facilitation, and suggested improvements for the workshop method. Following each workshop day, facilitators engaged in structured reflection on the session, including participant engagement, facilitation dynamics, and the implementation of workshop activities. Participant survey responses and facilitator reflections are reported in Appendix~\ref{sec:appendixa}.

\subsection{Data Analysis}
\begin{figure}[htbp!]
    \centering
    \includegraphics[width=1.0\columnwidth]{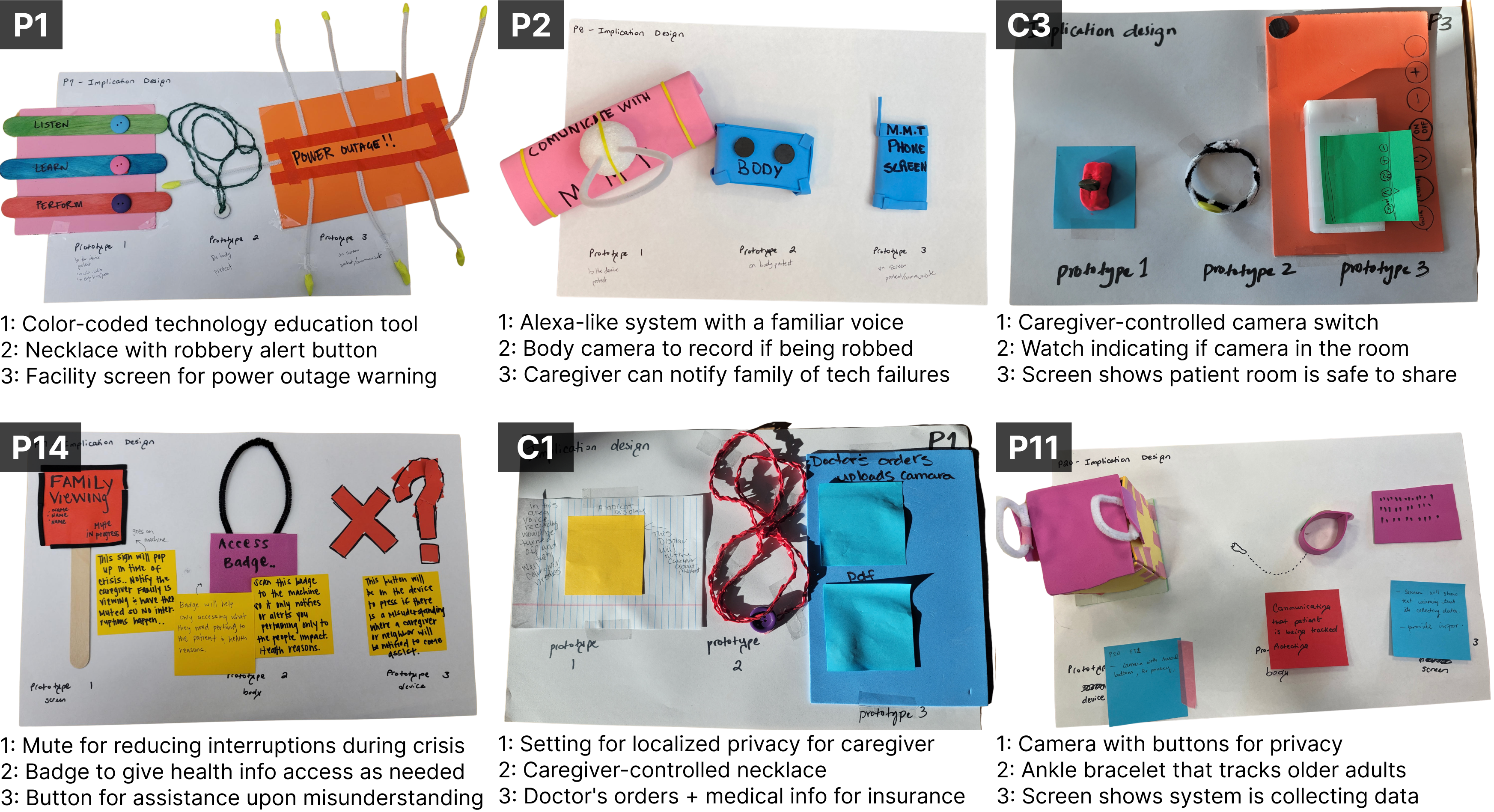}
    \caption{Six participant workstations from the Implication Design Round. Workstations had spots for all artifacts from this round of designing, giving flexibility to have multiple components and to add notes describing artifacts.}
    \label{fig:DataSheets}
\end{figure}

Participants documented their individual and group brainstorming, artifacts, and design descriptions on paper ``workstations'' (Figure~\ref{fig:DataSheets}). We applied thematic analysis~\cite{braun2006using} to approximately 12 hours of participant discussions and 156 generated artifacts and ideas. Audio was transcribed; Spanish-language recordings were translated into English and reviewed by a Spanish-speaking researcher. Occasional Tongan-language exchanges were not included in the analysis. Transcripts and videos were uploaded to HeyMarvin~\cite{HeyMarvin} for coding. The first author coded all six groups (36 hours of coding material), while two additional researchers independently coded two of six groups (12 hours), selecting groups they had not facilitated.

Following \citet{braun2006using}, we began with open coding of six hours of data to identify participants' descriptions of implications, values, and proposed modifications. Researchers then compared interpretations and consolidated an initial codebook, which was applied and iteratively refined across the remaining data. The final codebook comprised six domains, 27 parent codes, and 67 child codes and mapping of codes to findings are presented in Appendix~\ref{sec:appendixc}.

We identified value tensions by examining instances in which participants described competing desired outcomes, conflicting needs among affected parties, or situations in which mitigating one implication could produce another. We developed design recommendations by examining participants' proposed modifications alongside the implications they were intended to address. Artifacts were interpreted together with participants' verbal and written descriptions rather than independently inferring their meaning. This process converged on three areas corresponding to our research aims: perceived ethical implications of \textit{CSAI}, underlying value tensions, and participant-generated recommendations for redesigning \textit{CSAI} to better support professional care work.

\subsection{Researcher Positionality}
Our research team spans diverse racial, socioeconomic, and disciplinary backgrounds, including civil engineering, human-building interaction, computer science, HCI, and design, with research focused on human-centered infrastructure for aging in place, accessibility, health, and wellbeing. The first author and lead facilitator identifies as a Black woman from a low-income background with family ties to PHA and CNA work; other facilitators identify as Asian, European, white, and Latin American, reflecting some participant identities. None of the researchers identified as Tongan, creating cultural distance from PHA participants that may have influenced facilitation and interpretation. Our interest in whether \textit{CSAI} should exist and under what conditions led us to encourage open discussion. However, our academic and technical backgrounds may have privileged technological possibilities over care-work realities. 

\section{Findings}
We organize our findings around the research questions. We first present the ethical implications PHAs and CNAs anticipated from \textit{CSAI} (\textbf{RQ1}), followed by the value tensions they identified
(\textbf{RQ2}). We then present design guidelines derived from participants' proposed modifications to address these implications and tensions
(\textbf{RQ3}). Participants' self-reported backgrounds are shown in Table~\ref{tab:caregiving-background}. Throughout the findings, we use participants' terminology for older adults: PHAs referred to them as \textit{``clients,''} while CNAs used \textit{``patients.''}

\subsection{RQ1. Ethical Implications of \textit{CSAI}}
During the workshop, participants identified both positive and negative implications of \textit{CSAI} (see Table~\ref{tab:implications} in Appendix~\ref{sec:appendixa}) and developed prototypes that communicated or addressed potential harms while preserving perceived benefits (see Figure~\ref{fig:all}). Four implications emerged: (1) \textit{CSAI} creates different risks and benefits across care roles and settings; (2) integrating \textit{CSAI} creates new responsibilities for caregivers; (3) \textit{CSAI} may intensify competition between PHAs; and (4) PHAs fear \textit{CSAI} could reduce their job security. 

\subsubsection{\textbf{\textit{CSAI} creates different risks and benefits across care roles and settings}}
The workshops revealed contextual differences between CNA and PHA participants, which shaped their values and perceived implications. PHAs viewed themselves as integral to the care pipeline through their embodied knowledge of older adults’ day-to-day lives, yet expressed that this contribution was not fully recognized. As P1 explained, 

\begin{quote}
\textit{``Us caregivers, we have to face the challenges. The doctors and nurses don't come and put them [on] the toilet. We are the ones. I always [say] \textbf{we are the dirty hands}.''}
\end{quote}

PHAs viewed \textit{CSAI} as a technology that could significantly impact home caregivers and were, therefore, more willing to provide insights into implications. In contrast, CNAs described their role as primarily focused on monitoring older adults in skilled facilities and relaying observations to nurses, who hold formal responsibility for clinical decision-making. CNAs also spoke of how \textit{``in a facility [older adults are] watched 24/7... CNAs, nurses, even some of our housekeepers keep an eye out on them''} suggesting that \textit{CSAI} may not be necessary in skilled facilities due to the available support. 

Training and professional preparation further distinguished the two groups (Table \ref{tab:caregiving-background}). CNAs reported having institutional training through formal educational institutions, which provided them with standardized medical knowledge and credentials. P6 and P16, two PHAs, noted that they previously held CNA certifications but were working as PHAs due to their licenses being expired, which highlights the professional mobility of CNAs and their ability to transition across related care roles as needed. PHAs described more varied and informal pathways into care work. Many emphasized that their training was rooted in cultural expectations, family caregiving, and lived experience rather than formal certifications. As P8 explained, \textit{``As Tongans, we [are raised on] the island to take care of our parents. So when we moved here, majority of the women, that's what they do for a living, work as caregivers.'' } These findings reveal that variability in care work is institutionally structured, shaping how PHAs and CNAs differently position themselves and their labor, and therefore, communicate the implications of \textit{CSAI}. The findings also point to the need for AAL sensing systems that are contextually relevant to specific care roles, as \textit{CSAI} may impose different risks or offer different forms of support, depending on caregivers’ institutional positioning and care settings.

\begin{table*}[ht]
\centering
\caption{Self-reported caregiving background by participant ($N$=24).}
\label{tab:caregiving-background}
\scriptsize
\setlength{\tabcolsep}{3pt}
\renewcommand{\arraystretch}{0.9}

\begin{threeparttable}
\begin{tabular*}{\textwidth}{
@{\extracolsep{\fill}}
lll
lll
lll
}
\toprule

\multicolumn{3}{c}{\textbf{C-G1 (CNA, $n$=4)}} &
\multicolumn{3}{c}{\textbf{C-G2 (CNA, $n$=2)}} &
\multicolumn{3}{c}{\textbf{P-G1 (PHA, $n$=4)}} \\
\cmidrule(r){1-3}
\cmidrule(lr){4-6}
\cmidrule(l){7-9}

\textbf{ID} & \textbf{Experience} & \textbf{Training} &
\textbf{ID} & \textbf{Experience} & \textbf{Training} &
\textbf{ID} & \textbf{Experience} & \textbf{Training} \\

C1 & 5--6 yrs & Institutional &
C4 & $<$1 yr & Institutional &
P1 & $>$10 yrs & Institutional \\

C2 & 3--4 yrs & Institutional, Practical &
C6 & $>$10 yrs & Institutional &
P2 & 3--4 yrs & Online \\

C3 & 3--4 yrs & Practical &
& & &
P3 & 3--4 yrs & Declined \\

C5 & 9--10 yrs & Institutional &
& & &
P4 & 9--10 yrs & Declined \\

\midrule

\multicolumn{3}{c}{\textbf{P-G2 (PHA, $n$=5)}} &
\multicolumn{3}{c}{\textbf{P-G3 (PHA, $n$=4)}} &
\multicolumn{3}{c}{\textbf{P-G4 (PHA, $n$=5)}} \\
\cmidrule(r){1-3}
\cmidrule(lr){4-6}
\cmidrule(l){7-9}

\textbf{ID} & \textbf{Experience} & \textbf{Training} &
\textbf{ID} & \textbf{Experience} & \textbf{Training} &
\textbf{ID} & \textbf{Experience} & \textbf{Training} \\

P5 & $>$10 yrs & Declined &
P10 & $>$10 yrs & Online, Practical &
P14 & $>$10 yrs & Practical, Online \\

P6 & $>$10 yrs & Institutional, Practical &
P11 & 5--6 yrs & Institutional, Practical &
P15 & 7--8 yrs & Declined \\

P7 & 3--4 yrs & Practical &
P12 & $<$1 yr & Practical &
P16 & $>$10 yrs & Declined \\

P8 & 7 yrs & Practical &
P13 & $<$1 yr & Declined &
P17 & $>$10 yrs & Practical \\

P9 & $>$10 yrs & Online &
& & &
P18 & 1--2 yrs & Practical \\

\bottomrule
\end{tabular*}

\begin{tablenotes}
\scriptsize
\item Institutional = in-person collegiate/vocational;
Online = workplace/virtual;
Practical = on-job.
\end{tablenotes}

\end{threeparttable}
\end{table*}

\subsubsection{\textbf{Integrating \textit{CSAI} creates new responsibilities for caregivers}}\label{sec:protocols}
Participants voiced their perspectives on the potential challenges of integrating \textit{CSAI} in their respective work settings. The CNA workshop surfaced one main challenge with \textit{CSAI} being in skilled facilities: privacy restrictions. For example, CNAs highlighted that violations of HIPAA\footnote{Health Insurance Portability and Accountability Act (HIPAA) is a U.S. federal law that protects patients' health information \cite{assistance2003summary}} posed a great risk to facilities and their continued employment. Due to limitations in information access in facilities governed by HIPAA, CNAs expressed that \textit{CSAI} would be better suited for home settings. For example, C1 explained, 

\begin{quote}
\textit{``This [system] would be good for independent living and assisted living, not a skilled nursing facility. Because of their protocols... there can't be anything monitored. We can't even have video cameras in the facilities. You can't discuss [anything]. You can't even leave your computer open. Everything has to be locked all the time.''} 
\end{quote} 

To address these concerns, all participants prototyped security (Figure~\ref{fig:all}: \#70, \#71, \#72, \#73, \#140) and control features (Figure~\ref{fig:all}: \#67, \#69, \#92). There were additional views on the practicality of \textit{CSAI}. For example, C2 said, \textit{''depending on the condition of the patient, the system may not be good for all to use.'' } Most CNAs spoke of how patients with dementia or stress may not be able to engage with \textit{CSAI}. C1 also reflected on her previous experience using similar systems, noting that technology-supported processes created additional burden for her: \textit{``There [was] a lot of information you [had] to enter--50 questions per patient.''} These perspectives potentially speak to how AAL may significantly impact caregivers due to their close proximity to older adults. Because \textit{CSAI} is intended to rely heavily on multimodal interaction through user feedback, caregivers may be required to enter significant amounts of detailed information for system performance accuracy (e.g., particularly for older adults who may not be able to engage with the system), given their in-depth contextual knowledge gained through continuous observation.

The challenges with \textit{CSAI} in home settings, described by PHAs, primarily stemmed from resistance among older adult users. For example, P8 suggested that affluent older adult clients who hire caregivers specifically would prefer to hire them, rather than purchase technology to help them: \textit{``[For] the clients [who] are rich; that's why they pay us. They're not going to use those.''} Similarly, P1 noted that structured technologies may create routines that make older adults feel confined or institutionalized: \textit{``With routines, they feel like they’re prisoners. They [feel] institutionalized [like] they don’t have their right[s].''} Echoing these concerns, C-G1 emphasized that older adults' opposition to technology poses a significant barrier to adoption. P3 reinforced this point, sharing that she faces technology resistance from older adults on a daily basis: \textit{``Especially dealing with the old timers.'' } Likewise, P1 recounted previous experiences with clients who expressed discomfort and distrust toward such systems, recalling one who skeptically asked, \textit{``I’m just curious, what are you recording?''} PHAs further emphasized that they were responsible for ensuring the comfort and safety of older adults, noting that affecting a client’s wellbeing may result in client complaints and possible termination. For example, P1 highlighted how her inability to operate technology resulted in her client being burned and therefore blamed for it: 

\begin{quote}
\textit{``[I worked] in a facility that [had a] fancy toilet. You press [the button] to splash the water and the [toilet] warms up. [When] I press it, it was too hot. I didn't know how to operate it.''} 
\end{quote}

These findings suggest that caregivers will face additional responsibilities, including learning how AAL sensing systems operate, explaining their functionality to older adults, and supporting those who are skeptical of or resistant to the systems.

\subsubsection{\textbf{\textit{CSAI} may intensify competition between PHAs}}
PHAs noted that \textit{CSAI} could create potential competition between them, highlighting uncertainty of whether they or their care agencies could navigate this conflict successfully. P-G2 mentioned that if care agencies adopted \textit{CSAI}, they may charge clients higher service fees. As a result, they expressed that \textit{CSAI} could intensify competition between PHAs if some chose to bypass working through care agencies that use \textit{CSAI} to negotiate directly with clients to provide lower-cost, \textit{``under the table''} care. P5 shared her experience losing her job in a similar way:

\begin{quote}
\textit{``I was taking care of [a] couple. We lived in a hotel. [A] friend who [took] care of another client [spoke] to the [couple] and [told them they] can spend less money [for her to] take care of them. And of course, I lost my job.''} 
\end{quote}

These perspectives highlight a vulnerability in caregiving job security and the potential for technology to shift labor into a hidden workforce where formal protections are largely absent.

\subsubsection{\textbf{PHAs fear \textit{CSAI} could reduce their job security}}
Many PHAs expressed concern that \textit{CSAI} could impact their job security. For example, P13 said that a positive implication of \textit{CSAI} is that it reduces caregiver burden resulting in less stress and work but could \textit{``cut down our hours.''} P-G2 discussed their concern with being fired if \textit{CSAI} were implemented. While brainstorming, P5, unprompted, asked the facilitator, \textit{``Does [this system] take away my job?''} When the facilitator explained that the purpose of the workshop was to explore the ethical implications, P5 replied, \textit{``Disable the machine''} and \textit{``Kick the robot.''} P9 then posed the question, \textit{``What will disappear if the technology is successful?''} She then answered herself, saying, \textit{``[The] caregiver!''} P15 also mentioned destroying \textit{CSAI} in this context. These findings highlight strong opposition to AAL sensing systems among some home-based caregivers, who perceive these systems as potential threats to their livelihood. 

C-G1 also voiced that \textit{CSAI} could affect the employment of caregivers working with independent older adults. C1 asked, \textit{``If everything is smart, why would you need a caregiver?''} All members of C-G1 had the same question, with C3 stating that she had a personal conversation with C2 regarding \textit{CSAI} replacing caregivers. However, they concluded their jobs were relatively secure due to them working with clients who have high support needs and require human interaction. Although we envision \textit{CSAI} as a tool for independent living, the workshop format encouraged participants to speculate on how \textit{CSAI} might impact different contexts (e.g., settings), for which CNA participants generated a range of implications across homes, skilled care facilities, hospitals, gyms, and community centers. Despite engaging in this speculative exercise, CNAs ultimately expressed limited concern that their own jobs were at risk due to the nature of the older adults they care for. This contrast suggests that the workshop activities can surface insights about broader societal implications even when participants’ personal perceptions or concerns differ.

\subsection{RQ2. Value Tensions Around \textit{CSAI}}
Participants collectively navigated their value tensions by considering how \textit{CSAI} could make their work experiences visible and help navigate challenges with older adults and their families, while recognizing that the same capabilities could also expose behaviors that could be used against them.

\subsubsection{\textbf{Camera data can protect caregivers while exposing them to surveillance}}\label{sec:evidence}
Several PHAs stated that cameras could expose the negative experiences they face with older adults, such as cruelty, racism, and violence. P16 noted, \textit{``Sometimes the client will be rude. Then when we tell the family, they don't believe us. So we [can] go to the camera.''} P14 highlighted her similar experience with clients and how a camera being present could influence their behavior:

\begin{quote}
\textit{``The positive is the client will behave nicely if they are on the camera. Because sometimes, for example, [if] it's a white guy or he's racist, they pinch you, knock you, and hit you. It's a hard thing.''} 
\end{quote}

However, P16 and P18 emphasized that camera footage could be misinterpreted. They explained that viewers could impose their own biases when interpreting camera data, potentially harming the caregiver. They also mentioned that agencies or families might rely solely on the camera's perspective, ignoring relevant context. P5 prototyped a button for logging caregiver activities as supplementary evidence to complement camera footage (Figure~\ref{fig:all}: \#102). P11 further described how caregivers already experience scrutiny from family members, who may question how they provide care: \textit{``We don't like it sometimes because they try to text us, like, `Can you not do my mom like that [i.e., treat or handle my mom that way]?' We're doing our job.''} She proposed adding a sign to notify the caregiver when the family is watching. Aligned with some PHAs, C5 saw cameras in the home as a form of protection for caregivers, providing evidence of erratic client behavior when necessary. Yet as a CNA who had previously worked as a PHA, she also expressed a desire to protect her own privacy and avoid being recorded in the skilled facility, revealing a tension between surveillance as a protective mechanism and surveillance as a threat to worker privacy that varies by care setting.

\subsubsection{\textbf{Accountability can become evidence of caregiver ``neglect''}}
Some participants made note that \textit{CSAI} could use for accountability. P8 suggested that caregivers’ awareness of \textit{CSAI} monitoring them could instead function as an accountability tool and therefore modified \textit{CSAI} to send reminders to those \textit{``neglecting their duties''}, prompting corrective action before families are alerted (Figure~\ref{fig:all}: \#105). P10 added a similar feature to a mobile phone to remind PHAs to focus on their clients (Figure~\ref{fig:all}: \#129). P11 and P13 described using \textit{CSAI} for this type of caregiver surveillance as ``\textit{micromanaging},'' highlighting differences in value among PHAs. P5 similarly considered how \textit{CSAI} could provide evidence that benefits caregivers, such as documenting the number of hours worked to support requests for a raise (Figure~\ref{fig:all}: \#102).

However, participants also recognized that sensing capable of documenting their work could capture instances when caregivers were not providing expected care. One example P9 alluded to was caregivers falling asleep. She modified \textit{CSAI} to include a wakefulness activity for the caregiver and client (Figure~\ref{fig:all}: \#82). Another example focused on situations where PHAs would leave clients home alone. P8 shared, \textit{``[The family may] want to monitor to make sure if [they're] paying the caregiver, [that she is] not sneaking out at night to dance and come back.''} P6 added, \textit{``or sneaking out with the boyfriend.''} 

 Participants in P-G4, P14 and P17, prototyped on/off buttons to protect from being on camera (Figure~\ref{fig:all}: \#87 and \#90), while P16 added the ability to disconnect the camera (Figure~\ref{fig:all}: \#89). P-G4 collectively spoke of camera use making PHAs uncomfortable and impacting their behavior. For example, P14 stated that \textit{CSAI} would make caregivers \textit{``slaves to the camera.''} The group felt that caregivers behave honestly on their own and should not feel forced to behave in a certain way by camera presence. In the CNA workshop, C-G1 discussed the potential for cameras to catch ``neglect'' or negligent actions and noted the heightened risks for medical personnel if there was such evidence. C1 said, \textit{``If a caregiver, CNA, or nurse were found negligent, they could be sued.''} 

Concerns about evidence against PHAs also surfaced during the warm-up in reference to \textit{Toilet Buddy}. For example, several PHAs noted they were prohibited from using any bathrooms in client homes but sometimes did so anyway, fearing that \textit{Toilet Buddy} might expose this. Some suggested modifications, such as separate toilets for caregivers and clients (Figure~\ref{fig:all}: \#14) or features to conceal usage, like P7’s \textit{``Hulk''} destroying the toilet (Figure~\ref{fig:all}: \#12). Additionally, some PHAs worried \textit{Toilet Buddy} could reveal unauthorized guests, such as significant others. P5 proposed an alarm to notify of unwanted visitors (Figure~\ref{fig:all}: \#10). Both \textit{Toilet Buddy} (implied to not have a camera) and \textit{CSAI} (implied to include one) surfaced the capacity to disclose unauthorized caregiver behaviors, highlighting that equity in design cannot be achieved simply by adopting seemingly \textit{``less-intrusive’’} modalities. Overall, the findings suggest significant risk to caregiver wellbeing, as AAL sensing systems may increase surveillance in ways that expose perceived workplace negligence or unproductivity.

\subsubsection{\textbf{CSAI can challenge family authority while supporting caregiver expertise}}
While \textit{CSAI} could give families greater oversight of caregivers, some PHAs also envisioned it as a tool to help them navigate conflicts with families in a way that empowers them as caregivers. This was also reflected in features that enabled families to access information about the older adult (Figure~\ref{fig:all}: \#151, \#155, \& \#156). These views also prompted reflection on how the gaps in health awareness within older adults’ social communities may affect the quality of care. For example, P11 shared an instance where her client’s family friend misinformed her that the client had Alzheimer’s disease when she actually had schizophrenia and bipolar disorder. She highlighted how this misinformation led to care that triggered a negative reaction from the client, suggesting that care for Alzheimer’s disease (a neurodegenerative condition) differs from care for schizophrenia and bipolar disorder (psychiatric conditions). She further stated,

\begin{quote}
\textit{``[Families] have a fixed mindset about their loved one and see no issues with the care instructions they provide. However, the caretaker may experience a different reality upon arrival.''} 
\end{quote}

\subsubsection{\textbf{CSAI can support caregiver mediation of clients' family relationships}}
Beyond managing their own conflicts with families, some PHAs envisioned \textit{CSAI} as supporting their existing role in navigating relationships between older adults and their family members. P6 suggested using \textit{CSAI} to handle this dynamic by \textit{``set[ting] up times for clients to talk with their family.''} P9 responded, \textit{``What about if the client doesn't want to talk to the family?''} P6 then said, \textit{``You can always try, you know. [You say] your son is on the phone.''} P6’s statement demonstrates her active engagement in facilitating and managing client–family relationships. P4, who was not part of this group, similarly spoke to the caregiver’s role in family dynamics, stating, \textit{``Sometimes the clients will listen to the caregiver better than the family members,''}, highlighting caregivers' potential impact.

A few PHAs also discussed using \textit{CSAI} to manage conflicts that they directly witness, even if they are not personally involved. This mediating role extended beyond facilitating family connection to navigating conflicts within older adults' social relationships. P-G2 raised examples of older adult romantic partnerships in which one partner might need convincing to adopt \textit{CSAI}. P9 expressed concern that a significant other might oppose \textit{CSAI}, while P5 and P6 noted that a boyfriend might resist adopting the system in order to save money. P8 further suggested, ``\textit{Because [the boyfriend] might be sneaky, so he doesn't want something like that},'' implying that a significant other might resist \textit{CSAI} due to concerns of his behaviors being exposed. These findings suggest significant power dynamics between older adults and their families that PHAs may have to mediate, presenting opportunities to consider how \textit{CSAI} could facilitate this role, such as by supporting communication between older adults and their family members.

\subsection{RQ3. Modifications for Addressing Ethical Implications and Value Tensions}
Participants proposed ways to redesign \textit{CSAI} to mitigate potential harms while maintaining its benefits. In doing so, they identified concrete changes to better support both care and the work required to provide it.

\subsubsection{\textbf{Design system around care roles, settings, and responsibilities}}
PHAs saw \textit{CSAI} as a potential learning tool for caregivers; however, they repeatedly emphasized that any system designed to support care work well must be relevant to their day-to-day realities and align with their capabilities. 

\paragraph{\textbf{Integrating sensors for client safety and monitoring}} 
Participants highlighted how, in its current form, \textit{CSAI} was ill-equipped to support older adults and caregivers. Therefore, they requested sensors and features that generated data to support effective client monitoring. For example, PHAs suggested safety measures by adding sensors for 1) measuring water temperature for clients who struggle with water temperature awareness, 2) measuring room humidity for safety during showers, and 3) informing caregivers when the older adult is using a toilet for up-to-date monitoring. P1 stated that including measures of movement and activities is crucial, especially if clients have dementia, as they might run the bathtub, highlighting safety concerns. P8 also said that \textit{CSAI} ``should monitor the patient's balance'' for safety. PHAs prototyped many modifications to keep track of clients, such as a body camera on the older adult (Figure~\ref{fig:all}: \#99), a leash (Figure~\ref{fig:all}: \#101), and an ankle monitor (Figure~\ref{fig:all}: \#111). In support of the monitoring potential of \textit{CSAI}, P8 stated:

\begin{quote}
``\textit{Maybe they are sneaking and eating a bar of chocolate, and they didn't want you to know. You know, you're going to find out in the morning that the diabetes went up just because she ate a bar of chocolate. And it will be caught by this technology. That's a good thing.}''
\end{quote}

P15 wanted to add a vision system that could identify an older adult moving their head or legs, signaling they are trying to get out of bed. C3 also wanted a bed sensor to know when a patient is getting up. As CNAs have to access medical-related information, they proposed integrating sensors that could measure patient vitals such as \textit{``[blood] sugar, [blood] pressure, and [blood] oxygen [levels]''} (Figure~\ref{fig:all}: \#139). These findings suggest that AAL sensing configurations should be both context-aware and tailored to specific care roles, such as PHAs requiring enhanced monitoring systems for older adults living independently at home, and CNAs needing sensors that provide vital health updates for patients with high support needs. Regarding features, both PHAs and CNAs prototyped alerts to notify them when something was wrong with the older adult (Figure~\ref{fig:all}: \#109, \#124, \#137, and \#139) or to inform the caregiver when they are not following protocols (e.g., diet restrictions (Figure~\ref{fig:all}: \#138)). 

PHAs also spoke of \textit{CSAI} being integrated into other areas of older adults' life for continuous monitoring. P1 felt that extending the system outside of the home is important for when older adults go on a walk. P-G2 discussed that \textit{CSAI} should be designed to accompany the older adult in settings like dinners or vacations. Similar to this, some PHAs felt that for the system to be reliable it should always be operational, even when there is no power, so they could always access information about clients (Figure~\ref{fig:all}: \#122, \#123, and \#125). These ideas and artifacts highlight that the continuous monitoring and safe management of older adults are central responsibilities of caregivers, and point to how context-aware AAL sensing systems could meaningfully support this labor.

\paragraph{\textbf{Selecting the right sensors and placement for optimal client care}}
PHA participants further emphasized what to consider for sensor type and placement. P1 voiced the importance of sensor type, preferring a camera over the other sensors, as it would reveal exactly what clients are doing: \textit{``[With] the sensors, the caregiver knows the patient is inside, but with the camera, you're keeping track.''} She noted that this could allow caregivers to leave the room while still monitoring the older adult, providing older adults with greater independence and autonomy. P-G1 also suggested using silent sensors because \textit{``noises freak older adults out.''} P2 added, ``\textit{Sensor sounds can make the client confused or feel unsafe [and] frustrated.}'' P-G1 also had a discussion about designing for sensor placement with P1 saying to consider mobility and where older adults spend their time: ``\textit{most prefer the living room for its TV, comfort, and pride as the main area. Many enjoy relaxing in their Lazy Boy recliners.}'' As designers consider AAL sensing systems more broadly, there is a need to account for how these systems will be shaped by the built environment and how they can be integrated across different contexts. 

\begin{figure*}[!t]
  \centering
  \includegraphics[width=\textwidth]{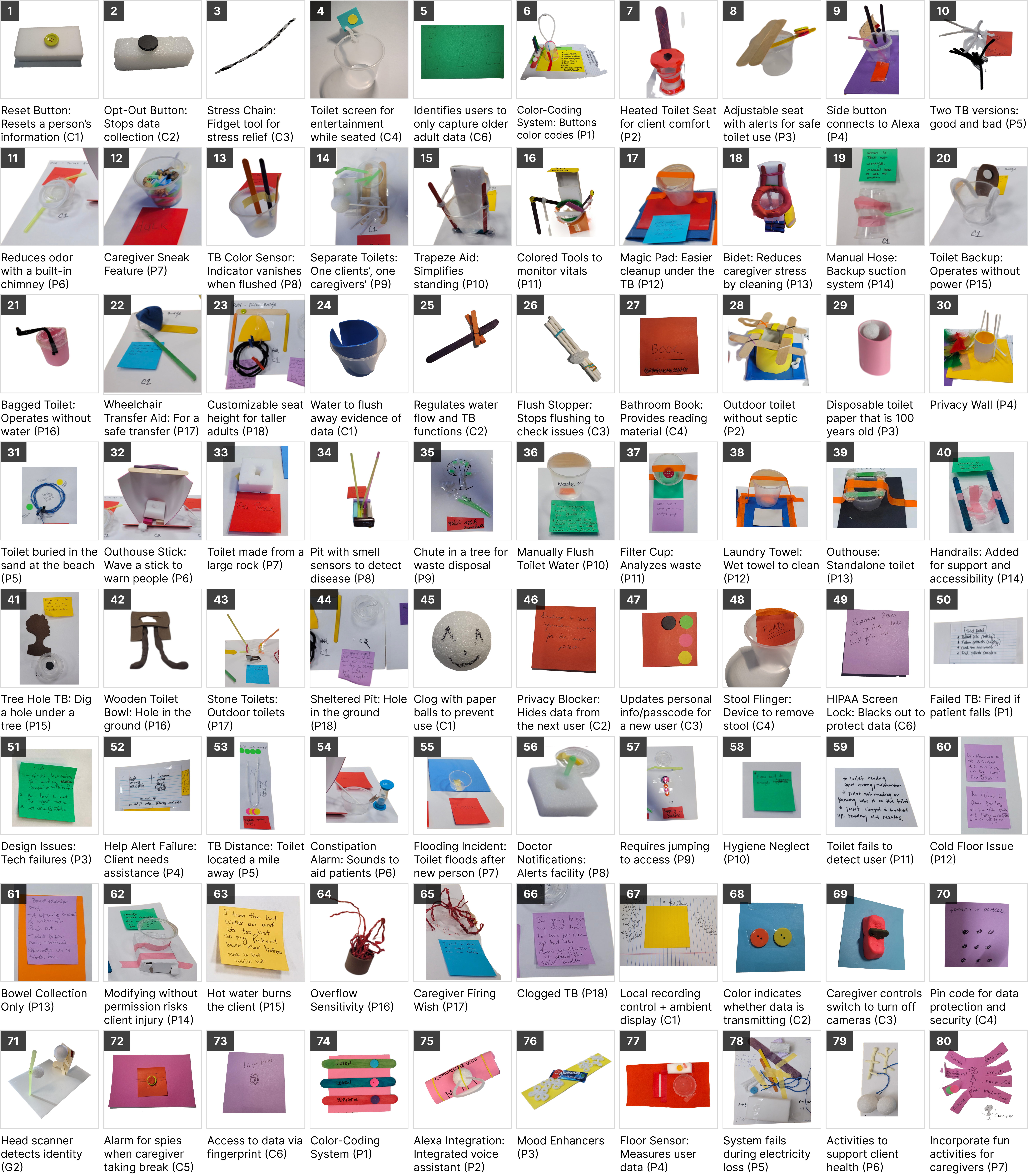}
  \caption{Participant artifacts with labels: Items 1--23 are Toilet Buddy modifications under constraint \#1 (use magic), 24--44 are modifications under constraint \#2 (use 100 year old technology), and 45--66 are modifications under constraint \#3 (get yourself fired).}
  \label{fig:all}
\end{figure*}

\begin{figure*}[!t]\ContinuedFloat
  \centering
  \includegraphics[width=\textwidth]{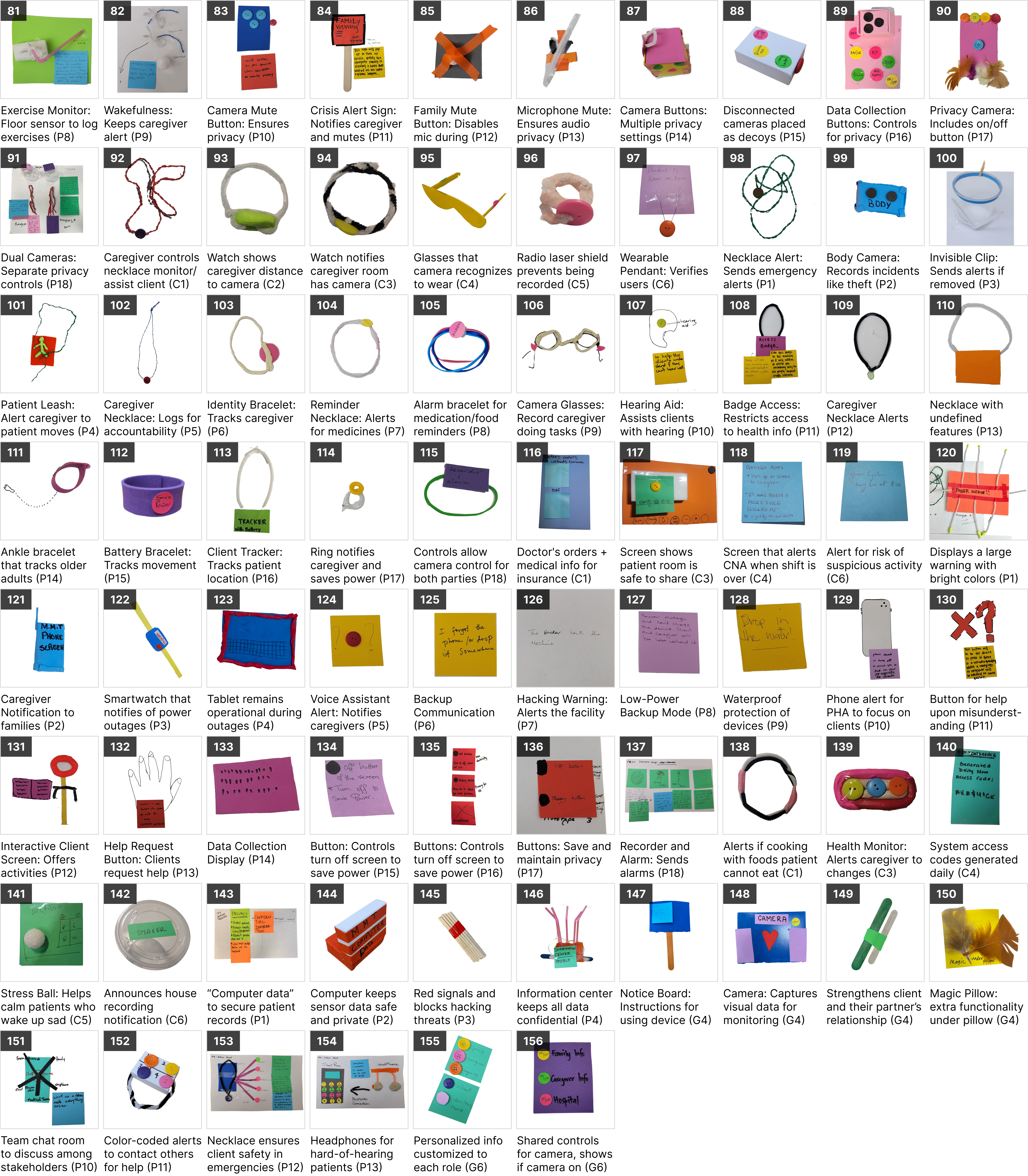}
  \caption{(Cont’d) Participant artifacts with labels: Items 67--91, 92--115, and 116--137 are modifications to the device, the body, and the screen, respectively, and were created during the Implication Design Round. Items 138--156 are modifications created during the Action Round.}
  \label{fig:all}
\end{figure*}

\subsubsection{\textbf{Support career development without replacing hands-on and interpersonal training}}
There was an opposing view on \textit{CSAI}'s ability to appropriately upskill PHAs. C1, who was in school for a nursing degree, expressed that \textit{CSAI} was not an effective tool to provide the information needed to upskill to a CNA. For example, in regard to the exam prospective CNAs take for licensing, she noted that the system was only sufficient for the written portion of the exam but not the hands-on skills portion. She added, 

\begin{quote}
    \textit{``I don't think this would help to train a CNA. Because to train a CNA, you need a lot of hands-on. It's a hands-on approach. I have to train someone to be a CNA next week. [There] are skills that they teach you when you're working on the floor in a skilled nursing facility that you would only learn if you did your training on the floor.''} 
\end{quote}

The CNA exam is typically administered in two parts: a written or oral knowledge assessment and a hands-on skills assessment~\cite{registerednursing2025cna}. The exam is state-administered with requirements varying by state~\cite{registerednursing2025cna}. C1 added a result of the implication and provided examples of the learning needs: 

\begin{quote}
\textit{``It's your first day, right? You passed everything, you got your job, but you've never had a hands-on approach with the person. You're going to be scared out of your mind}...[They] would need to observe skin breakdown\footnote{Skin breakdown, a form of tissue damage caused by sustained pressure, is highly prevalent among older adults~\cite{medline_skin_breakdown_prevention_2024}} and assess the coloring of the skin.''
\end{quote}

She noted that in addition to these medical-related skills, interpersonal skills are also taught ``\textit{on the floor}'' (i.e., working directly in the real care environment with patients) and would need to be included in \textit{CSAI}. This concern highlights the critical need for collaborative design with care administrators and other relevant parties to ensure that training requirements are properly integrated and that technologies align with real-world practices.

\subsubsection{\textbf{Give caregivers control over how they are sensed and observed}}
Despite concerns of surveillance, some participants highlighted that cameras were beneficial for monitoring older adults. Several participants expressed interest in using a passive camera to subtly influence client behavior but modified it to reduce surveillance through controls to manage camera features. P15 suggested \textit{``placebo''} cameras that were turned off but installed everywhere (Figure~\ref{fig:all}: \#88) so the client would behave, but the caregiver would not be recorded. This design choice highlights the complex tensions around surveillance and the need for features that shift behavioral responsibility onto older adults rather than caregivers. Similarly, P10 did not want the client's family to see or hear her through a camera. However, having the camera was beneficial for her own protection, so she suggested managing at least the listening capacity, using mute buttons as a solution (Figure~\ref{fig:all}: \#83). P18 suggested a two-camera system, one to capture the caregiver and another to capture the client, with both having different privacy settings (Figure~\ref{fig:all}: \#91). CNA participants, C1, C3, and C5 also added controls to turn cameras off only in rooms caregivers are in (Figure~\ref{fig:all}: \#67, \#69, and \#72) to obtain privacy and maintain visibility of older adults.

\subsubsection{\textbf{Support family coordination without displacing caregiver or older-adult autonomy}}
PHAs prototyped features that integrate family into client care (Figure~\ref{fig:all}: \#153, \#156). P-G3 suggested that family integration could help reduce micromanagement of caregivers' methods, using \textit{CSAI} as evidence for and validation of their approach. To address these possibilities, P-G2 modified \textit{CSAI} to take on the role of encouraging a significant other to adopt the system by integrating features that attended to romantic relationships. They framed these as: 1) \textit{``voodoo''} to convince the husband (Figure~\ref{fig:all}: \#150), 2) a tool to strengthen relationships (Figure~\ref{fig:all}: \#149), and 3) a camera that would encourage spending time together (Figure~\ref{fig:all}: \#148). Overall, these findings suggest that AAL sensing systems that support interventions (e.g., personalized care for relational needs) that address older adults' broader dimensions of wellbeing needs (e.g., social wellbeing) can potentially help foster environments in which caregivers are less likely to manage or be subjected to occupational harm such as client-family conflict.

\section{Discussion} 
This study examined how PHAs and CNAs perceive \textit{CSAI}. We structure our discussion around our three research questions: (1) ethical implications for professional caregivers; (2) negotiating values in ambient assisted living; and (3) centering professional caregivers in AAL design, which includes caregiver-informed design guidelines for AAL sensing systems. PHA participants positioned themselves as integral to the care community, describing themselves as the ``dirty hands'' who perform much of the hands-on work that makes aging in place possible. CNA participants positioned themselves as having a more limited role within the care team, primarily monitoring older adults and relaying observations to nurses who held greater responsibility for clinical decision-making. This study is significant as one of the few to use participatory design to engage a largely overlooked population---professional caregivers---in redesigning and proposing modifications to AAL that center the needs of their respective communities.
 
\subsection{\textbf{Ethical Implications for Professional Caregivers}}
Our findings show that AAL sensing systems will not enter a neutral home but a workplace already shaped by unequal access to resources, protections, and authority. AAL can redistribute responsibility, visibility, and care work within the home while leaving caregivers with limited power over how these changes impact their work and wellbeing. 

\subsubsection{\textbf{Care infrastructure shapes caregiver responsibility}}
Having two types of caregivers working in different institutional settings allowed us to see how existing infrastructural conditions shape the implications of \textit{CSAI}. Our findings showed that the responsibilities and consequences associated with the system may depend on the training, resources, and authority available to caregivers in their respective care settings. For CNAs, institutional protocols (e.g., HIPAA) and defined role boundaries shape how they interact with technology and provide care, while also giving them clearer expectations about their responsibilities and greater peace of mind about their job security. PHAs, in contrast, described greater flexibility in how they provided care in older adults' homes, but this flexibility came with fewer institutional protections and greater uncertainty about losing their jobs.

\paragraph{\textbf{Care infrastructure determines who becomes responsible}}
These differences are particularly important when considering who becomes responsible for managing AAL. The same information from \textit{CSAI} may create different expectations for a PHA working alone in an older adult's home and a CNA working within a facility. A CNA may have an established protocol, supervisor, or care team to turn to when the system identifies a concern, whereas a PHA may be left to determine whether and how to respond. Designing around care roles therefore requires considering whether caregivers have the training, authority, and support needed to act on the information they receive. Prior work has raised concerns that home care workers may be held responsible for problems that emerge from AI systems~\cite{solano2025running}; our findings show how this concern may extend to AAL, as participants anticipated being blamed for problems involving technologies they may not understand or be empowered to manage. If \textit{CSAI} identifies a change in an older adult's condition, for example, receiving that information should not automatically make the caregiver responsible for determining its clinical significance. Instead, the system could clarify what action falls within the caregiver's role and when the concern should be escalated to another member of the care team. Participants' button and alert prototypes for different parties, such as family members, caregivers, and doctors, point toward this kind of role-based personalization. Such designs could help prevent information from becoming caregiver responsibility simply because the caregiver is the person physically present.

\paragraph{\textbf{AAL can produce new forms of ``care work''}}
At the same time, participants anticipated that integrating \textit{CSAI} into care could create new forms of work. Our findings suggest that designing AAL that provide caregivers with additional information can also create new expectations for what they should know, monitor, or act on. Caregivers may be expected to use AAL, learn how the system operates, and help older adults use it. These responsibilities extend beyond direct care and could introduce additional ``care work,'' such as technical work, into an already demanding role. One PHA's experience with an unfamiliar ``fancy toilet'' illustrates this concern; when the technology did not work as expected and the older adult was harmed, the caregiver described being blamed despite not knowing how the system operated. Similar situations could arise with AAL if caregivers are expected to integrate them into care work.

\paragraph{\textbf{AAL should support caregiver recognition}}
If operating and interpreting AAL becomes part of care work, caregivers should be recognized and compensated for this work. Participants discussed using \textit{CSAI} to provide evidence of the care they perform and support requests for raises, indicating a desire for AAL use to translate into higher pay and other direct benefits for caregivers. AAL may also introduce new technical skills into care work, creating opportunities for these skills to be recognized as part of caregivers' professional development. This recognition could include connecting experience with AAL to formal training, credentials, or advancement opportunities. Participants, however, also identified limits to what \textit{CSAI} could currently contribute to professional development. One CNA felt that the system could not support career advancement because it lacked the capacity to teach caregivers about physiological factors relevant to care. This finding suggests that using AAL as a pathway for professional development would require more substantive learning support. For example, per our participants, systems could help caregivers understand health conditions, recognize changes in an older adult's status, or learn appropriate care responses. For PHAs who may move from home care into skilled facilities, such support could help make the skills developed through AAL use more relevant to advancement across care settings.

\subsubsection{\textbf{AAL can reshape access to care work}}
We also found that \textit{CSAI} could interact with the existing precarity in access to care work. PHAs often work part-time, earn low wages, and rely on public assistance~\cite{campbell2021caring,phi2019ushome}. Participants worried that \textit{CSAI} could further reduce their paid hours or replace them as the system takes on aspects of their work. They also emphasized that older adults may avoid technologies they do not want to use or pay for, potentially shifting care work away from agencies and toward caregivers working ``under the table.'' This shift may remove some of the agency oversight and support available to PHAs, as digital systems may unintentionally reconfigure labor toward less protected forms of work~\cite{himani2025transformative,ming2023go}. The result could leave PHAs with fewer protections when navigating the mistreatment and abuse participants described. These issues create an important distinction between reducing caregiver burden and reducing caregiver work. While AAL may reduce responsibilities such as continuous monitoring, for PHAs whose income depends on providing this labor, reducing the amount of care work required may also reduce their paid hours or access to work. AAL could shape both who or what provides care and how care work is distributed (e.g., between caregivers and technology). 

\subsection{\textbf{Negotiating Values in Ambient Assisted Living}}
The tensions show how the same features of \textit{CSAI} can create different consequences for care work. Making hidden aspects of care work visible can support caregivers while also exposing them to greater scrutiny. However, using \textit{CSAI} to manage relationships between older adults and their families can relieve some relational work while changing the role caregivers play in those relationships.

\subsubsection{\textbf{Making care work visible can support and expose caregivers}}
Prior work draws attention to the invisible work performed by home health aides, showing how important aspects of care often go unrecognized, including relational labor with family members that also surfaced in our findings~\cite{ming2023go}. AAL sensing systems may make a largely hidden workforce significantly more visible. This creates an opportunity to make the work and working conditions of professional caregivers more visible within the broader care landscape but also raises questions about what this new visibility will mean for workers who have historically received limited visibility~\cite{ming2023go}. However, the concern is not visibility itself, but how information about caregivers moves across the home care context. Information produced in the course of providing care may take on different meanings when captured by \textit{CSAI} and made available to families, agencies, or others. Thus, the value of visibility depends not only on what becomes visible but also on who has access to that information, for what purpose, and how it is interpreted or used~\cite{nissenbaum2019contextual}. Our participants noted that this visibility could work in caregivers' favor, such as by encouraging older adults to modify harmful behavior, documenting mistreatment or abuse, or providing evidence that supports caregivers' accounts of what occurred. A caregiver-centered \textit{CSAI} could therefore allow caregivers to preserve and report evidence of an incident when they need support. Such features could be particularly important for caregivers with limited workplace protections~\cite{campbell2021caring,phi2019ushome}, as such documentation could provide evidence of working conditions that might otherwise be difficult to substantiate. 

Caregiver-controlled records of repeated aggression or other unsafe conditions could potentially support requests for organizational changes or broader advocacy around working conditions. However, another finding shows how the same visibility could work against caregivers. PHAs also described \textit{CSAI} as a system that could identify when they were not doing their job. While this visibility could support accountability, it could also give families or agencies greater ability to evaluate caregivers' work without capturing the context surrounding their actions. Our participants noted that caregivers could face employment or legal consequences based on misinterpretation. \citet{ming2025exploring} raises questions about who worker data benefits and how those data are used. Our findings extend this tension to AAL, where \textit{CSAI} could produce new forms of evidence about caregivers without giving them equivalent power over its interpretation or use. This issue creates an asymmetry in which families or agencies with power over caregivers' employment may gain greater visibility into their work without caregivers gaining a corresponding ability to contest or contextualize how that information is interpreted. Addressing this asymmetry requires considering how caregivers can provide context for recorded behavior, aligning with prior work that argues workers should be able to contextualize camera footage~\cite{chowdhary2023can}. This tension creates a technical opportunity for AAL to provide caregivers with accessible ways to know when they are being monitored, limit monitoring when appropriate, and add context to information that may otherwise be misinterpreted.

\subsubsection{\textbf{Supporting care relationships requires more than supporting individual parties}}
Participants surfaced the work they do to support relationships between older adults and their families, known as relational labor~\cite{ming2023go}. Families may experience uncertainty about an older adult's health condition, which can result in them putting pressure and stress on caregivers~\cite{karlsson2019home,fitzwater2000violence}. Our participants noted that \textit{CSAI} could potentially relieve some of this work; however, addressing these tensions may require more than providing separate information or controls for each party (e.g., older adults, families, and caregivers) in the system as suggested by the prototypes. Per our findings, this could include ways for \textit{CSAI} to bring older adults and their families together. Designing AAL for care therefore requires considering the needs of older adults, families, and caregivers, as well as the relational labor caregivers perform between them. This finding extends HCI conversations about relational care by showing how technologies introduced into care relationships may become part of how caregivers facilitate communication, navigate tensions, and maintain relationships between other parties (e.g., through proof of care or supporting documentation).

However, our CNA workshop also showed that relational work is more complex than a burden to be reduced for home caregivers. A CNA participant described how developing interpersonal skills and experience is crucial for PHAs seeking to advance in their careers. Interpersonal skills, including building relationships with older adults and their families, are competencies in direct care~\cite{campbell2021caring}. This issue complicates the role of AAL sensing systems such as \textit{CSAI} in PHA career advancement, as relieving relational work may also reduce opportunities to develop these skills. For example, following a difficult interaction with a family member, the system could offer optional resources related to communication,
de-escalation, or advocating for an older adult's preferences. Future work should further explore how AAL can relieve relational work without limiting opportunities for caregivers to develop the interpersonal skills needed for career advancement.

\subsection{\textbf{Centering Professional Caregivers in AAL Design}}
Our findings show that designing AAL sensing systems requires accounting for the people who perform care work alongside them, not just the older adults these systems are intended to support. 

\subsubsection{\textbf{Caregivers should shape the role and boundaries of AAL}}
Participants highlighted the extent of the precarity of home care work by showing how caregivers may lack access to basic resources, such as bathrooms in clients' homes, experience physical mistreatment such as being pinched, and receive incorrect information from families about older adults' health conditions. They also highlighted power imbalances in their relationships with older adults and their families, such as clients having economic power to hire them and family members assuming power through their questioning of how they perform their work. These existing conditions matter for HCI because technologies introduced can interact with and potentially intensify them. Centering mainly older adults and informal caregivers in conversations about AAL therefore risks designing systems that improve care outcomes while overlooking, or even intensifying, harms experienced by professional care workers. 

Collective speculation is valuable for enabling communities impacted by emerging technologies to explore their benefits, harms, and potential alternatives~\cite{haghighi2023workshop}. As surfaced in our findings, AAL has the potential to shape not only individual experiences but also the broader community within the workforce and care infrastructure. We extend human-centered work by considering how technologies interact with community norms and structures and how design can account for these broader levels. Our findings also show the richness of knowledge professional caregivers bring as contributors to AAL design~\cite{bartle2023machine}. Much AAL work focuses on recognizing changes in behavior and activity for predictive capabilities~\cite{vimarlund2021ambient}. Drawing from their experiences providing care, our participants proposed modifications that reflected the work required to maintain aging in place and the conditions under which that work occurs. At the same time, their modifications considered older adults' independence, privacy, and autonomy, including reducing the need for caregivers to remain in the same spaces as older adults. These findings show how caregivers' knowledge of everyday care can inform AAL design in ways that account for their own work while continuing to support the goals of aging in place.

Some participants also rejected \textit{CSAI} or questioned whether sensing should be involved in their work at all. Thus, caregivers' design knowledge was not limited to proposing new features but also included determining what \textit{CSAI} should support and when the technology should not be used. Caregivers' role in AAL design should extend beyond providing requirements for a predetermined system. If professional caregivers are expected to work alongside AAL systems and experience the consequences of their use, they should also have influence over what these systems do and where their involvement in care should end.

\subsubsection{\textbf{Caregiver-Informed Design Guidelines for Ambient Assisted Living Sensing Systems}}
We translate caregivers' anticipated implications, value tensions, and proposed modifications into five design principles (Table \ref{tab:designguidelines}) for expanding the design AAL sensing systems for professional caregivers: (1) configure AAL systems around distinct care roles and settings; (2) provide caregivers with control over monitoring and incident reporting; (3) design AAL systems to mediate relational tensions; (4) integrate context-aware sensing in locations where it supports care; and (5) support professional learning without reducing care expertise.  

\begin{table*}[!t]
\centering
\caption{Design guidelines for caregiver-centered AAL sensing systems}
\label{tab:designguidelines}
\small
\renewcommand{\arraystretch}{1.25}
\setlength{\tabcolsep}{6pt}

\begin{tabularx}{\textwidth}{
    >{\raggedright\arraybackslash}p{0.24\textwidth}
    >{\raggedright\arraybackslash}X
    >{\raggedright\arraybackslash}p{0.31\textwidth}
}
\toprule
\textbf{Design Guideline} &
\textbf{Why It Matters} &
\textbf{Example Design Considerations} \\
\midrule

\textbf{Configure AAL around distinct care roles and settings}
&
Caregivers differ in their responsibilities, training, authority, resources, and institutional constraints. These differences shape what information they need and what actions they can take.
&
Role-based information and alerts; setting-specific data collection; responsibilities matched to caregiver authority. \\[2pt]

\textbf{Provide caregivers control over monitoring and incident reporting}
&
Caregivers need meaningful ways to understand and respond to monitoring, particularly when resulting data may be used to evaluate their work~\cite{bartle2023machine,chowdhary2023can}.
&
Monitoring indicators; viewing notifications; mechanisms to contextualize records or report incidents; agreed-upon controls for disabling sensing. \\[2pt]

\textbf{Design AAL to mediate relational tensions}
&
AAL operates within relationships among older adults, caregivers, families, and other care professionals~\cite{harnett2024use}. Differences in authority, knowledge, and preferences shape how information affects these relationships.
&
Role-based information sharing; communication preferences; escalation pathways; safeguards against unnecessary oversight. \\[2pt]

\textbf{Integrate context-aware sensing where it supports care}
&
Sensing should provide actionable information about activities and risks relevant to care rather than maximize data collection. What is useful depends on older adults' routines, mobility, and living environments.
&
Caregiver-informed sensor placement; sensing adapted to routines and mobility; collection limited to information relevant to care. \\[2pt]

\textbf{Support professional learning without reducing care expertise}
&
Professional development should complement rather than replace the experiential and interpersonal dimensions of caregiving.
&
Training on physiological changes, communication, protocols, and regulations; connections to hands-on training and career advancement. \\

\bottomrule
\end{tabularx}
\end{table*}

\section{Limitations and Future Work}
We conducted a participatory design study with CNAs and PHAs. Given the qualitative nature and limited sample size, our findings may not generalize to broader CNA and PHA populations. Our sample also included a larger representation of Pacific Islander PHAs, which may reflect experiences specific to this demographic. Participants' limited design experience and language barriers may have also shaped participation, though we do not expect these limitations to affect the rigor of the study. Because participants considered the impacts of \textit{CSAI} on themselves, our findings are largely caregiver-centered. They are also more heavily shaped by PHA perspectives, as our sample included approximately three times as many PHAs as CNAs, reflecting our intention to amplify PHA voices while using CNA perspectives to contextualize caregiver career needs. Future work should examine additional caregiver perspectives alongside those of older adults to better understand these intersections in AAL design.

\section{Conclusion}
We examined the ethical implications and value tensions of \textit{CSAI}, an AAL sensing system, from the perspectives of CNAs and PHAs. Our findings show how these systems can reshape caregivers' responsibilities, visibility, and care work within already precarious care environments. Participants also identified how \textit{CSAI} could support their work while introducing new risks and where its role in care should end. Based on these findings, we propose caregiver-informed design guidelines that account for distinct care roles and settings, provide caregivers greater control over monitoring, and recognize their role in shaping the boundaries of AAL in care.

\begin{acks}
We would like to thank many individuals for their contributions to this paper. First, we express deep gratitude to the caregivers for participating in this research. Second, we thank Dr. Julie N. Thai, Dr. Can Chen, Dr. Haya Rubin, and Salote Tupou for their assistance with recruitment. Next, we thank Alacia Johnson and Kenya Bodley for their support in developing the concept video. Finally, we appreciate all feedback from the Stanford HPDS/HAI lab, Jianna So, and Helena Vasconcelos. This work was supported by a grant through the Stanford Institute for Human-Centered Artifical Intelligence and funding from the Stanford Graduate Fellowship.
\end{acks}

\bibliographystyle{ACM-Reference-Format}
\bibliography{main}

\appendix

\section{Additional results}\label{sec:appendixa}

\subsection{Results from the Anticipation Round}

\begin{table}[!htbp]
\centering
\caption{The context, use cases, and direct stakeholders identified by each group during the Anticipation Round}
\label{tab:usecase}

\footnotesize
\setlength{\tabcolsep}{4pt}
\renewcommand{\arraystretch}{1.2}

\begin{tabularx}{\linewidth}{
    @{}
    >{\raggedright\arraybackslash}p{0.11\linewidth}
    >{\raggedright\arraybackslash}p{0.13\linewidth}
    >{\raggedright\arraybackslash}X
    >{\raggedright\arraybackslash}p{0.19\linewidth}
    >{\raggedright\arraybackslash}p{0.23\linewidth}
    @{}
}
\toprule
\textbf{Group} &
\textbf{Context} &
\textbf{Use Case} &
\textbf{Direct Stakeholders} &
\textbf{Indirect Stakeholders} \\
\midrule

\multicolumn{5}{@{}l}{\textbf{Workshop 1 (CNA, $n=6$)}} \\
\midrule

C-G1 (C1--C3, C5\textsuperscript{*}) &
Bathroom &
For caregivers who want to monitor incontinence, the system can detect and measure waste amount via bathroom sensors. &
Caregivers, older adults with gastrointestinal issues &
Doctors, family members, researchers \\
\midrule

C-G2 (C4, C6) &
Mobile application &
For nurses and family members, the app allows input on care decisions. &
Spouses, children, authorized healthcare personnel &
Insurance companies, power of attorney \\
\midrule

\multicolumn{5}{@{}l}{\textbf{Workshop 2 (PHA, $n=18$)}} \\
\midrule

P-G1 (P1--P4) &
Bathroom &
For older adults or caregivers focused on safety, the floor sensor monitors activities. &
Older adult, caregiver(s), family members, neighbors, close friends &
Health professionals, family members, caregivers, nurses, doctors, first responders, occupational therapist, cleaners, delivery person, security system, physical therapist, neighbors \\
\midrule

P-G2 (P5--P9) &
Older adult's home &
For caregivers who want clients to live a long life, the system provides alerts/recommendations on nutrition and hydration. &
Doctors, friends, family, children, nurse(s)/aides, caregivers, neighbors, clients &
Insurance, grandchildren, lawyers, hospitals, church communities \\
\midrule

P-G3 (P10--P13) &
Older adult's home &
For caretakers, the app provides crisis alerts based on home sensors to enhance safety. &
Older adults, caregivers, family, nurses, doctors, neighbors &
Home health aides, nurse advisers, close friends, trusted relatives \\
\midrule

P-G4 (P14--P18) &
Rehab facility &
For families, caregivers, and older adults, the camera provides safety monitoring in rehab. &
Caregiver, client, on-site individuals &
Rehab agency, client's family, visitors, fire department \\
\bottomrule

\end{tabularx}

\vspace{3pt}
\begin{minipage}{\linewidth}
\footnotesize
\textit{Note.} \textsuperscript{*}C5 did not attend Day 1 and joined C-G1 on Day 2.
\end{minipage}

\end{table}

\onecolumn
\begin{longtable}[t!]{p{0.10\textwidth} p{0.40\textwidth} p{0.40\textwidth}}
\caption{The ethical implications identified by each group during the Anticipation Round}%
\label{tab:implications}\\
\toprule
\textbf{Group} & \textbf{Positive Implications} & \textbf{Negative Implications}\\
\midrule
\endfirsthead

\endhead

\midrule
\multicolumn{3}{r}{\textit{Continued on the next page}}\\
\endfoot

\bottomrule
\endlastfoot

C-G1 &
\begin{itemize}[leftmargin=2em]
    \item Centralized system for paper-based records
    \item More efficient patient monitoring
    \item Upskilling for CNA written exam
    \item Identifying side effects of drugs
    \item Proof of bowel movements if patient is disoriented
    \item Improved drug research and medication
    \item Enables doctors to identify patterns
    \item Useful for research on drug absorption and disease data
\end{itemize} &
\begin{itemize}[leftmargin=2em]
    \item Overuse of laxatives due to data insights
    \item May feel overwhelming for nurses unfamiliar with technology
    \item Potential for over-worrying and frequent doctor visits
    \item Possible radiation concerns with excessive technology
    \item Could disrupt caregiver workflow
    \item Eligibility requirements add work; the system is not suited for all
    \item May not fully prepare for hands-on PHA upskilling
    \item Risk to caregiver jobs (redundancy concern)
\end{itemize}\\
\midrule

C-G2 &
\begin{itemize}[leftmargin=2em]
    \item Facilitates health tracking and risk assessment
    \item Easily accessible and shareable health info
    \item Reduces healthcare and emergency resource strain
    \item Preventative measures for serious events (e.g., heart attack)
    \item Supports under-served populations
    \item Enables rapid emergency responses
    \item Allows friends to give health feedback
\end{itemize} &
\begin{itemize}[leftmargin=2em]
    \item Risk of unauthorized access by friends/others
    \item Privacy concerns if accessed by burglars
    \item Potential for stress from information overload
    \item Risk of confusion from tracking multiple users
    \item Reduced separation between work and personal life for the caregiver
    \item Risk of hacking and data breaches
    \item App information may become outdated
\end{itemize}\\
\midrule

P-G1 &
\begin{itemize}[leftmargin=2em]
    \item Reduces the risk of falls
    \item Improves individual health
    \item Enhances quality of caregiver's work
    \item Provides instant, worldwide access
    \item Increases safety for patients and caregivers
    \item Simplifies caregiving tasks
\end{itemize} &
\begin{itemize}[leftmargin=2em]
    \item Causes confusion for patients unfamiliar with technology
    \item Increases risk of theft
    \item Systems shut down during power outages
    \item May compromise patient privacy
    \item Limits personal freedom
    \item Increases stress if technology malfunctions
\end{itemize}\\
\midrule

P-G2 &
\begin{itemize}[leftmargin=2em]
    \item Supports client health
    \item Reminds caregiver of client needs
    \item Increases client independence
    \item Assists caregiver with scheduling
    \item Improves communication
    \item Alerts family if caregiver is absent
\end{itemize} &
\begin{itemize}[leftmargin=2em]
    \item May create unsafe situations
    \item Risk of caregiver job loss
    \item Reduces relaxing environment between client and caregiver
    \item May lead to competition between caregivers
    \item Could result in caregivers becoming overly dependent on the system
\end{itemize}\\
\midrule

P-G3 &
\begin{itemize}[leftmargin=2em]
    \item Offers guidance on effective caregiving
    \item Notifies family or caregivers in emergencies
    \item Assists in reminding clients of tasks
\end{itemize} &
\begin{itemize}[leftmargin=2em]
    \item Can lead to family micromanaging caregivers
    \item May distract caregivers
    \item Misunderstanding instructions may arise
\end{itemize}\\
\midrule

P-G4 &
\begin{itemize}[leftmargin=2em]
    \item Protects caregivers with cameras (e.g., racist attacks)
    \item Ensures respectful behavior from clients
    \item Makes caregiving tasks easier
    \item Provides evidence of incidents for agencies
\end{itemize} &
\begin{itemize}[leftmargin=2em]
    \item Increases stress for older adults
    \item Invades caregiver privacy
    \item Forces caregivers to act a certain way
    \item Potential dependency on electricity
    \item High costs involved
\end{itemize}\\

\end{longtable}

\subsection{Perceptions of collective speculation workshops}\label{sec:perceptions}
We report participant and facilitator reflections on the adapted workshop, including participants' responses to participation, challenges encountered during participation, understanding and engagement, and considerations for using collective speculation with similar communities.

\subsubsection{\textbf{Participant perceptions of the workshops}}\label{sec:participantperceptions}

\paragraph{\textbf{Responses to participation}}
Most participants responded favorably to this workshop. On a Likert scale from 1 (poor) to 10 (excellent), participants rated the workshop an average of 9.95 ($n$=21). In this section, we further highlight their perceptions of participating in the workshop and any barriers they mentioned associated with their ability to participate. Participants of both workshops expressed their appreciation for the design aspects of the workshop. C2 stated, \textit{``[I liked] to imagine the difference of creating something that would help in working with patients.''} C4 said, \textit{`` [I liked the thought-provoking questions that challenged me to think differently than the norm.''} C6 added, \textit{``I liked sharing our thoughts and brainstorming.''} Other shows enthusiasm for working in groups. For example, C1 said, \textit{``[I] liked the grouping. [I] didn't feel singled out.''} P17 added, \textit{''[I liked the] interaction [and] working in groups.''} P16 also said, \textit{``[I liked the] communication among the group.''} 

There was a positive response regarding the learning opportunity of the workshop. P18 stated, \textit{``[I liked] learning about the possibilities of advancing [older adults'] care with technology.''} P11 said, \textit{``[I liked] knowing what's to come and [giving] our input.''} Finally, P14 said, \textit{``[I liked] improving my ideas to be a good caregiver.''} Also, some participants showed appreciation for the facilitation. C5 said, \textit{``I liked how they explain and they have a lot of patience.''} P1 added, \textit{``Facilitators are excellent in guiding the workshop, explaining, and teaching.''} There were also positive acknowledgments expressed during the quieter moments of the workshop. P3 shared her excitement, saying, \textit{``This is fun. I haven't had that much fun for ages now. We bring out our inner child.''} Similarly, P1 emphasized the value of the workshop, stating, \textit{``I think it's a good idea to do a workshop like this with people like us who have experiences.''}

\paragraph{\textbf{Challenges with participation}}\label{sec:workshopchallenges}
A few PHAs shared that they needed more support from facilitators. For example, P15 said, \textit{``[I needed] more explanation,''} with P18 agreeing. There were also a few PHAs who were uncomfortable with the topics that were brought up. P12 stated, \textit{``[I didn't like] the idea of kicking the caregiver out,''} alluding to the system replacing her position. P11 didn't like the idea of \textit{``making a toilet with 100 years ago [technology].''} There were also expressions of the cognitive challenges they experienced. P8 stated, \textit{``This is really making you think.''} While C1 said, \textit{``I'm always a black and white thinker, so today I had to think outside of the box.''}

Additionally, most of the CNAs and many of the PHAs had expressed that they worked for several hours prior to participating in the workshop. They also made note that their participation was dependent on their ability to secure an overtime shift at work. 

\subsubsection{\textbf{Facilitator reflections of conducting the workshops}}\label{sec:facilitatorperceptions}

\paragraph{\textbf{Understanding and engagement}}
Facilitators reflected on challenges encountered during the workshops, particularly in explaining complex concepts, which often required simplifying instructions or relying on participants with higher English proficiency to support others. This dynamic sometimes led participants to take on informal ``co-facilitator'' roles, which, while beneficial, also raised concerns about potential power imbalances within groups. 

Time management was another difficulty, as some participants were unable to complete all activities due to the additional time needed for contextual explanations. Concepts such as ``implication design'' proved especially difficult to grasp, requiring additional time to explain and therefore prompting adjustments to the workshop structure, including the use of sticky notes to create low-fidelity sketches rather than full prototypes. Facilitators also struggled to balance guiding discussions with allowing participants space to develop ideas independently. Engagement levels varied across groups, with enthusiasm generally higher when participants were familiar with one another, though signs of fatigue and disengagement emerged toward the end of the sessions due to the workshops’ intensity.

\paragraph{\textbf{Workshop structure recommendations}}
Based on our experience, we propose the following recommendations for facilitators conducting collective speculation workshops with similar communities: 

\begin{enumerate}
    \item \textbf{Use accessible language:} Avoid design and field-specific jargon by simplifying design terms while preserving the core concepts and goals of the workshop.

     \item \textbf{Community-centered design:} For a community-centered approach, design scenarios that connect participants' collective identity to the technologies.

     \item \textbf{Structure for clarity and engagement:} Design guiding slides to reduce confusion and support smooth participation by using consistent visual cues for materials, visible timers to manage pacing, and color-coded sticky notes to clarify task purposes.

    \item \textbf{Select relevant constraints:} Consider selecting design constraints that align with both the introduced technology and the participants’ backgrounds. For instance, requiring the use of century-old technology could slow down idea generation for an intelligent toilet within a limited timeframe, or prove challenging for a group that is less comfortable with technology. 

    \item \textbf{Provide background and purpose:} Include introductory guiding slides that provide participants with clear background information on the purpose of the workshop and how its outcomes will be used, ensuring they understand their role and the overall objectives without shaping or biasing their input.

     \item \textbf{Leverage participants’ expertise early:} Begin with scenarios that draw on participants’ lived experiences to deepen discussions about the emerging technology of interest. For example, the relevance of the \textit{Toilet Buddy} scenario to caregiving helped participants think broadly about \textit{CSAI} during the Anticipation Round. 

    \item \textbf{Start with collaborative activities:} Early group exercises help build rapport and give facilitators insight into group dynamics. Make changes to groups as needed before the workshop begins. 
    
    \item \textbf{Review activities before timed exercises:} Allocate time to ensure participants understand tasks before starting timed sections.
    
    \item \textbf{Support independent exploration:} Allow participants to brainstorm and design without immediate intervention. Step in only when progress has clearly stalled to avoid hindering creativity or losing time.
\end{enumerate}

\section{Codebook and thematic coding}\label{sec:appendixc}

\begin{table}[H]
\centering
\caption{Qualitative codebook domains, definitions, and codes. Parent codes are shown in bold, followed by their child codes.}
\label{tab:domains}
\small
\renewcommand{\arraystretch}{1.15}

\begin{tabularx}{\textwidth}{
    >{\raggedright\arraybackslash}p{0.14\textwidth}
    >{\raggedright\arraybackslash}p{0.25\textwidth}
    >{\raggedright\arraybackslash}X}
\toprule
\textbf{Domain} & \textbf{Definition} & \textbf{Parent and child codes} \\
\midrule

Caregiver &
Caregivers, their roles, experiences, responsibilities, and wellbeing &
\textbf{Control}: older adult; system; lack; agency

\textbf{Accountability}: caregiver; support network; other

\textbf{Wellbeing} \\

\midrule

Infrastructure &
Environmental, organizational, and structural conditions surrounding care work &
\textbf{Support networks}: family; agency; health team; friends/neighbors/etc.

\textbf{Utilities}: water; electricity; internet; other

\textbf{Location}: home; facility; outside; spatial arrangement; other \\

\midrule

Older Adult &
Older adults' experiences, needs, care, and interactions with caregivers and technology &
\textbf{Wellbeing}

\textbf{Surveillance}: proof of actions; monitoring for care

\textbf{Diversity/inclusion}: disability; age; race; other

\textbf{Care}: physical; emotional; social; ethical; upskilling/learning; caregiver burden; efficiency \\

\midrule

Technology &
Features, capabilities, data practices, and design characteristics of technology &
\textbf{Modality}: voice; camera; app; phone

\textbf{Features}: notifications; reminders; other

\textbf{Innovation}: design choice; new idea; manual; other

\textbf{Data}: type; sharing; exploration \\

\midrule

Challenges \& Opportunities &
Factors affecting the implementation, use, risks, or opportunities of the system &
\textbf{Cost}: affordability; impact; question/inquiry

\textbf{Dynamics}: relationships; career/work; norms; culture; older adult; caregiver

\textbf{Safety}: risk/concern/harm; job/employment; other

\textbf{Encryption/security}: data; user; other

\textbf{Identification}: direct stakeholder; indirect stakeholder

\textbf{Tracking}: spying \\

\midrule

Ideology &
Broader perspectives, assumptions, and orientations toward technology and its implications &
\textbf{Views on technology}: tech is fine; tech is bad; privacy

\textbf{-isms} (i.e., social views such as racism, sexism, and ageism)

\textbf{Direction of perspective}: negative; neutral; positive \\

\bottomrule
\end{tabularx}
\end{table}

\begin{longtable}{p{0.32\textwidth} p{0.62\textwidth}}
\caption{Mapping of themes to codes from thematic analysis}
\label{tab:themes-codes}\\

\toprule
Theme & \textbf{Parent code}: child codes \\
\midrule
\endfirsthead

\multicolumn{2}{l}{\textit{Table \thetable{} continued}}\\
\toprule
Theme & \textbf{Parent code}: child codes \\
\midrule
\endhead

\midrule
\multicolumn{2}{r}{\textit{Continued on next page}}\\
\endfoot

\bottomrule
\endlastfoot


\multicolumn{2}{l}{\textbf{RQ1: Ethical Implications}} \\
\midrule

\textit{CSAI} creates different risks and benefits across care roles and settings
&
\textbf{Accountability}: caregiver;
\textbf{Wellbeing};
\textbf{Care}: caregiver burden;
\textbf{Dynamics}: norms; culture; caregiver
\\
\midrule

Integrating \textit{CSAI} creates new responsibilities for caregivers
&
\textbf{Wellbeing};
\textbf{Location}: home; facility;
\textbf{Diversity/inclusion}: age;
\textbf{Care}: physical; ethical;
\textbf{Dynamics}: career/work; norms;
\textbf{Encryption/security}: data;
\textbf{Safety}: risk/concern/harm;
\textbf{Tracking}: spying;
\textbf{Views on technology}: privacy
\\
\midrule

\textit{CSAI} may intensify competition between PHAs
&
\textbf{Wellbeing};
\textbf{Dynamics}: caregiver;
\textbf{Safety}: job/employment;
\textbf{Cost}: impact
\\
\midrule

PHAs fear \textit{CSAI} could reduce their job security
&
\textbf{Care}: efficiency;
\textbf{Dynamics}: career/work;
\textbf{Safety}: job/employment
\\
\midrule


\multicolumn{2}{l}{\textbf{RQ2: Value Tensions}} \\
\midrule

Camera data can protect caregivers while exposing them to surveillance
&
\textbf{Wellbeing};
\textbf{Support networks}: family;
\textbf{Surveillance}: proof of actions;
\textbf{Care}: caregiver burden;
\textbf{Modality}: camera;
\textbf{Dynamics}: career/work; older adult;
\textbf{Safety}: other; job/employment;
\textbf{Tracking}: spying;
\textbf{-isms}
\\
\midrule

Accountability can become evidence of caregiver ``neglect''
&
\textbf{Control}: lack;
\textbf{Accountability}: caregiver;
\textbf{Wellbeing};
\textbf{Support networks}: family;
\textbf{Surveillance}: proof of actions;
\textbf{Features}: notifications;
\textbf{Innovation}: system failure; design choice;
\textbf{Data}: sharing;
\textbf{Dynamics}: caregiver;
\textbf{Safety}: job/employment;
\textbf{Tracking}: spying
\\
\midrule

\textit{CSAI} can challenge family authority while supporting caregiver expertise
&
\textbf{Support networks}: family;
\textbf{Wellbeing};
\textbf{Dynamics}: norms; relationships
\\
\midrule

\textit{CSAI} can support caregiver mediation of clients' family relationships
&
\textbf{Wellbeing};
\textbf{Support networks}: family;
\textbf{Care}: caregiver burden;
\textbf{Surveillance}: proof of actions;
\textbf{Dynamics}: career/work; relationships
\\
\midrule


\multicolumn{2}{l}{\textbf{RQ3: Design Guidelines}} \\
\midrule

Design system around care roles, settings, and responsibilities
&
\textbf{Control}: older adult;
\textbf{Location}: home; spatial arrangement;
\textbf{Wellbeing};
\textbf{Surveillance}: proof of actions; monitoring for care;
\textbf{Care}: physical; efficiency;
\textbf{Features}: other;
\textbf{Innovation}: design choice; system failure;
\textbf{Data}: collection method;
\textbf{Dynamics}: older adult;
\textbf{Views on technology}: privacy
\\
\midrule

Support career development without replacing hands-on and interpersonal training
&
\textbf{Wellbeing};
\textbf{Care}: upskilling/learning; efficiency;
\textbf{Features}: other;
\textbf{Innovation}: design choice;
\textbf{Dynamics}: career/work
\\
\midrule

Give caregivers control over how they are sensed and observed
&
\textbf{Control}: agency; system;
\textbf{Support networks}: family;
\textbf{Surveillance}: proof of actions;
\textbf{Modality}: camera; voice;
\textbf{Features}: other;
\textbf{Innovation}: new idea; manual; design choice; system failure;
\textbf{Dynamics}: norms;
\textbf{Tracking}: spying;
\textbf{Views on technology}: privacy
\\
\midrule

Support family coordination without displacing caregiver or older-adult autonomy
&
\textbf{Care}: social; emotional;
\textbf{Dynamics}: relationships; older adult; caregiver;
\textbf{Identification}: indirect stakeholder
\\

\end{longtable}


\end{document}